\documentclass{elsarticle}
\usepackage{amsmath}
\usepackage{amssymb}
\usepackage{graphicx}
\usepackage{multirow}
\usepackage{ragged2e}
\usepackage{xcolor}
\usepackage{comment}
\usepackage{url}
\usepackage{tikz}
\usepackage{array}
\newcolumntype{?}{!{\vrule width 1.1pt}}

\begin{document}
\begin{frontmatter}
\title{A Data-driven Understanding of COVID-19 Dynamics Using Sequential Genetic Algorithm Based Probabilistic Cellular Automata}
\author[add1]{Sayantari Ghosh}
\ead{sayantari.ghosh@phy.nitdgp.ac.in}
\author[add2]{Saumik Bhattacharya$^*$}
\ead{saumik@ece.iitkgp.ac.in}

\address[add1]{Department of Physics, National Institute of Technology Durgapur, India}
\address[add2]{Department of E \& ECE, Indian Institute of Technology Kharagpur, India }
\cortext[cor1]{Corresponding author}

\begin{abstract}
COVID-19 pandemic is severely impacting the lives of billions across the globe. Even after taking massive protective measures like nation-wide lockdowns, discontinuation of international flight services, rigorous testing etc., the infection spreading is still growing steadily, causing thousands of deaths and serious socio-economic crisis. Thus, the identification of the major factors of this infection spreading dynamics is becoming crucial to minimize impact and lifetime of COVID-19 and any future pandemic. In this work, a probabilistic cellular automata based method has been employed to model the infection dynamics for a significant number of different countries. This study proposes that for an accurate data-driven modeling of this infection spread, cellular automata provides an excellent platform, with a sequential genetic algorithm for efficiently estimating the parameters of the dynamics. To the best of our knowledge, this is the first attempt to understand and interpret COVID-19 data using optimized cellular automata, through genetic algorithm. It has been demonstrated that the proposed methodology can be flexible and robust at the same time, and can be used to model the daily active cases, total number of infected people and total death cases through systematic parameter estimation. Elaborate analyses for COVID-19 statistics of forty countries from different continents have been performed, with markedly divergent time evolution of the infection spreading because of demographic and socioeconomic factors. The substantial predictive power of this model has been established with conclusions on the key players in this pandemic dynamics. 
\end{abstract}
\end{frontmatter}

\section{Introduction} 
\begin{figure*}
\begin{center}
\includegraphics[width=\linewidth]{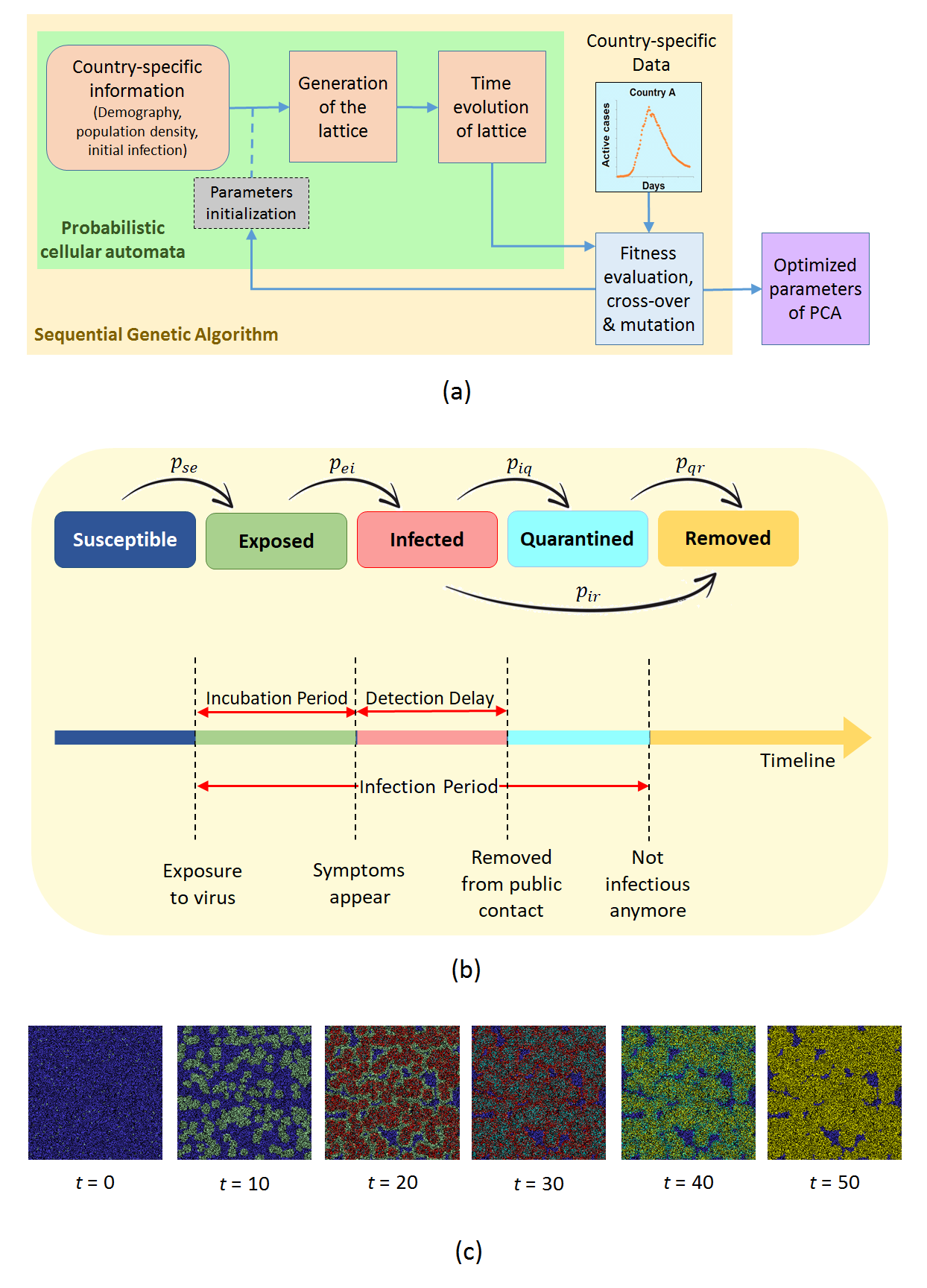}
\caption{An overview of the dynamics: (a) Object process diagram of the proposed model; (b) The schematic diagram of the disease transmission dynamics in form of a modified SEIQR model. Transition probabilities $p_{se}$, $p_{ei}$, $p_{iq}$, $p_{ir}$ and $p_{qr}$ are pointed out. The associated state transition delays are indicated on the timeline of the disease dynamics. (c) Time evolution of the spatial  lattice during spread of the infection in a population. The colors of the respective subpopulations, (i.e., susceptible, exposed, infected, quarantined and removed) are same as depicted in (a). }
\label{fig:scheme}
\end{center}
\end{figure*}

With its outbreak in Wuhan, China, Coronavirus disease-2019 (COVID-19) has spread across the world within a few months. Due to its explosive growth and considerable rate of fatality, World Health Organization (WHO) declared COVID-19 as a pandemic and a global health emergency \cite{who2020}. According to the available statistics in June, 2020, the total number of infections by SARS-CoV-2 (Severe Acute Respiratory Syndrome Coronavirus 2), the causative agent of this disease, is approaching 19 million around the world, causing around 700,000 deaths in 213 countries and territories, with no effective vaccination available in the market so far. Beyond respiratory discomforts including  pneumonia, dry cough, cold and sneezing \cite{jin2020epidemiological,pan2020clinical}, it has been reported to cause liver and gastrointestinal tract maladies, kidney dysfunction and heart inflammation, in cases of severe infection \cite{cheng2020kidney,han2020digestive,zheng2020covid}. This highly infectious disease transmits from person-to-person through respiratory droplets produced by infected person. Fomite-mediated and nosocomially acquired infections are also being identified as important sources of viral diffusion  \cite{wang2020immediate,wang2020unique,van2020aerosol}. A typical incubation time from exposure to symptoms has been reported for COVID-19, while infection transmission from asymptomatic individuals has been observed as well \cite{bai2020presumed,nishiura2020estimation,yu2020familial}. \\
Immediately after the detection of human-to-human transmission, the government agencies of various countries started implementing several mitigation strategies to control the epidemic. The measures thus taken include social distancing, restrictions on domestic as well as international travel, cancelling social events, shutting down of public as well as commercial activities etc. which can effectively reduce the possibilities of physical human contact. Moreover, contact tracing, aggressive testing as well as hospital or home quarantine for infected individuals and suspected cases have also been executed to track and prevent further spread. However, these strategies are directly contributing to enormous economical loss. The optimum estimation of this novel disease dynamics is emerging out as a challenging problem in this context. The immense disruption caused by COVID-19, resulting into overwhelming disorder in the health, economy and lives of billions of people around the globe, has brought the necessity for accurate modelling of infectious diseases into the focus. The effect and effectiveness of this complex interplay between differing length-scales and time-scales with the applied control strategies can only be understood and predicted with the help of precisely designed quantitative models. 
\subsection{Models for understanding COVID-19 statistics}
\noindent With a tremendous effort from researchers around the world, a spectrum of various mathematical and computational approaches is being used to understand and predict COVID-19 statistics, addressing its different perspectives. On a rudimentary sense, the studies being pursued can be segmented in two categories: (i) data science and machine learning approaches and (ii) differential equation based mathematical modelling techniques. The first group of studies trusted mostly on data mining from national/international repositories (e.g., WHO, country specific data centres etc.) or popular social media platforms to forecast the active cases and mortality data \cite{giordano2020modelling, yang2020modified,volpert2020quarantine,li2020retrospective,li2020propagation}. The major goal of these studies are to estimate and predict the time evolution of the disease using specific computational concepts, like Monte Carlo decision making, fuzzy rule induction, deep learning etc \cite{fong2020composite,chatterjee2020healthcare,fong2020finding,baltas2020monte,khatua2020fuzzy}. Some of these studies also explored impact of disease control interventions, like, travel restrictions \cite{liu2020covid}, patient quarantining and isolation \cite{traini2020modelling}, medical facilities \cite{lai2020assessing}, social distancing and administrative responsibility \cite{volpert2020quarantine} on epidemic spreading rate. Though these models are quite effective, being entirely dependent on data, the efficiency of these studies can be heavily inclined towards the data quality. As comprehensively reviewed by \cite{wynants2020prediction}, several data-dependent models are prone to suffer from high risk of bias, which is very much probable for imprecise short time series data.\\
With the evidence of giving effective predictions for past pandemics \cite{bauch2005dynamically,shinde2020forecasting, althouse2012synchrony}, the traditional approaches of the mathematical theory of epidemiological dynamics also have driven several researchers to study COVID-19 dynamics.  Theoretical modelling based approaches have been long associated to understand and predict the outbreak probabilities and seriousness of a disease, and provide key information to control the intensity \cite{anderson1992infectious,hethcote1973asymptotic,behncke2000optimal,bhattacharya2019viral}. Most of the mathematical models that are being used to investigate the COVID-19 dynamics \cite{liu2020reproductive,shim2020transmission,kucharski2020early,peng2020epidemic} are based on variants of classical deterministic model of susceptible-infectious-recovered (SIR) that was introduced by Kermack and McKendrick \cite{kermack1927contribution}. Constituting a set of nonlinear ordinary differential equations (ODE), the SIR model compartmentalises the population where susceptible subpopulation declines over time, constantly getting infected (by infectious subpopulation), and then recovered from (and gaining immunity to) the disease over time. Being powerful and computationally favourable tool to analyse epidemic, variants of this methodology are common in understanding real epidemic data \cite{rachah2017analysis,berge2017simple}. Though these models capture the disease transmission dynamics, being deterministic, they suffer from the assumption of homogeneous mixing, forgoing the spatial information.\\ For modelling real-world dynamics of a disease that spreads from close-contacts only, the tool needs to accommodate neighborhood information. Moreover, the platform requires to take into account of stochasticity of real dynamics, spatial infection spread and inherent heterogeneity in population, which are some major limitations of the mentioned works. Thus, the identification of research gap points out in a direction of designing a methodology that addresses the above mentioned issues to understand and predict neighbourhood-dependent person-to-person probabilistic transmission of COVID-19, that should be powered with extensive computational tools for parametric optimization.

\begin{table}
\centering
{
\caption{Comparison of the proposed method with the state-of-the-art COVID-19 models}
\label{comp}
\begin{tabular}{?l?l|l?} 
\hline
\begin{tabular}[c]{@{}l@{}}\textbf{\;\;\;\;\;\;Basic}\\\textbf{Methodology} \end{tabular} & {Differential equation models}                                                                                                                                             & {Data science~approaches}                                                                                                                                                            \\ 
\hline
\textbf{\;\;References}        &      \;\;\;\;\;\;\;\;\;\;\;\;\;\;\;\;\;\;[33-37], [39], [40]                                                                                                                                                                    &   \;\;\;\;\;\;\;\;\;\;\;\;  [13-22]                                                                                                                                                                                         \\ 
\hline
\begin{tabular}[c]{@{}l@{}}\textbf{\;\;\;\;\;\;}\\ \textbf{Limitations}\end{tabular} & \begin{tabular}[c]{@{}l@{}}a)  Homogeneous Mixing\\b) Most models are considered\\ as deterministic\end{tabular}                                                                                                      & \begin{tabular}[c]{@{}l@{}}\\a) No way to track person\\ to person transmission.\\b) No neighborhood\\ consideration.\end{tabular}                                                                                                                         \\ 
\hline
\textbf{Contribution}           & \multicolumn{2}{l?}{\begin{tabular}[c]{@{}l@{}}Proposed method,\\a) accommodates heterogeneity in population\\b) includes stochasticity and probabilistic dynamics\\c) estimates optimum epidemic dynamics parameters.\\d) considers neighborhood~and demography explicitly.\\e) performs robust prediction~with limited data.\end{tabular}}  \\
\hline
\end{tabular}
}
\end{table}
\subsection{Motivation and Contributions}
In this study, we propose probabilistic cellular automata based dynamical model, optimised through sequential genetic algorithm for an accurate assessment of the extent of COVID-19 dynamics. The major motivation of using cellular automata (CA) is its ability in depicting extremely complex macroscopic outcomes, while being based on local interactions that trusts on the interaction of a multitude of single individuals \cite{toffoli1987cellular,wolfram2018cellular}. This methodology is capable of giving a direct correspondence to the physical system and also rectifies the major drawbacks of ODE models by (i) tracking individual contact processes, (ii) giving room for introducing probabilistic individual behaviour, and (iii) capturing neighbourhood as well as global spatial information. Because of these reasons, CA based approaches have been successfully used as a competent substitute method to simulate physical, biological, environmental and social contagion-like spreading \cite{boccara1994probabilistic,beauchemin2005simple,fuks2001individual,willox2003epidemic}. For studying past epidemics as well as interpreting COVID-19, some studies have proposed cellular automata as an alternative method \cite{eosina2016cellular,pokkuluri2020novel,dascalu2020enhanced, ghosh2020computational}. However, to capture and interpret the behaviour of real data through CA needs a large-scale parameter optimization that could be time consuming as well as sub-optimal. Thus, though being extremely flexible and powerful, CA has not been yet optimized to understand and interpret COVID-19 data for countries worldwide. To explore this, in this study, genetic algorithm (GA) has been employed, which is a well-known method for generating the optimal parameter subset through stochastic search procedures based on the principle of the survival of the fittest \cite{wright1991genetic,yao1994nonlinear,katare2004hybrid,gulsen1995genetic,karr1995least}. Cross-over and mutations, two key properties of genetic algorithm help to optimize the parameter set efficiently in limited steps. Cellular automata coupled with genetic algorithm has been used before to explore evolutionary aspects of game theoretical problems \cite{schimit2016evolutionary}, but to the best of our knowledge analyzing and developing understanding from real pandemic data like COVID-19 using optimized CA platform has not been attempted yet. The main contributions of this work are as follows:
\begin{itemize}
    \item To build a CA model which is probabilistic, so that it can take into account of demographic variations, neighbourhood diversity and uncertainties of real dynamics.
    \item To create an easily implementable framework where optimization using GA will be done sequentially for all parameters associated with the transition rules of the CA model for real data interpretation.
    \item To interpret and understand COVID-19 disease transmission dynamics with an optimized CA framework, which can be extended for prediction as well.
\end{itemize}
Through this, on one hand, one can track the individual contact process through time and space; on the other hand, a self-adapting process of evolutionary strategies has been created by designing the chromosome with parametric genes and establishing fitness function that maximises over the generations. The main limitations of the state-of-the-art algorithms and the major contributions of the proposed method are listed in Table \ref{comp} for a clear understanding. {The main rationality behind this approach is that it is extremely difficult to find the optimal parameter of the complex spatial epidemiological model using random search or analytical techniques. The proposed GA based framework helps to search the parameter space more efficiently for the optimal performance of the entire algorithm}.\\
The rest of this article is organized as follows: Section 2 includes the proposed concepts of epidemiological model, probabilistic cellular automata and the sequential genetic algorithm used in this work. In Section 3, the results has been elaborately discussed where the optimized CA model has been employed for simultaneously understanding as well as analyzing active infections, total infections and total death caused by COVID-19 for several countries, considering the demographic and spatial population density variations. Section 4 is comprised of concluding remarks.
\section{Proposed Methodology}
{An object process diagram of the proposed method has been depicted in Fig. \ref{fig:scheme}(a). The methodology starts with the infection spreads following the SEIQR epidemiological model in a random human population over a 2D grid, initialized on a country-specific basis. The parameters of the epidemiological model is continuously optimized using proposed sequential genetic algorithm to match the real country-specific infection spread data.} The proposed methodology is consisted of three distinct parts$-$ (\textit{A}) epidemiological model that governs the infection spreading, (\textit{B}) probabilistic cellular automata (PCA) to model the dynamics of the pandemic spread and (\textit{C}) optimization of the parameters associated with PCA using genetic algorithm (GA) to fit real-world data. 
\subsection{Epidemiological Model}\label{ssec:subpop}
In the epidemiological model, the entire population is partitioned in five distinct parts. At the very beginning, every person was healthy but they are vulnerable to the infection. These people are denoted as susceptible ($S$) subpopulation. At time instance $t=0$, some people in the population got exposed to the infection from some known or unknown source. These exposed people do not have any particular symptom of the infection, but they can spread the infection to the susceptible people. These asymptomatic people are referred as exposed ($E$) subpopulation.  At time instance $t=0$, there were also some people who had clear symptoms of the infection and they also had the potential to spread the infection among susceptible people. This symptomatic people are considered as infected ($I$) subpopulation. After an incubation period, some of the exposed people show the symptoms of the infection and they move to subpopulation $I$. Because of the health facilities and testing time, the infected people are detected with some average delay, and put to quarantine. The people who are quarantined cannot spread the infection to other people, though they themselves remain in the infectious stage. These people are denoted as quarantined ($Q$) subpopulation. Both the quarantined people and the infected (but not detected) people would come out of the infectious stage eventually, and after that they no longer contribute in the infection spreading dynamics. 
\begin{table*}[tbp]
\centering
{{
\caption{Descriptions of the parameters used in the proposed work}
\label{params}
\begin{tabular}{|c|l|}
\hline
\textbf{Notation}                                    & \multicolumn{1}{c|}{\textbf{Description}}                                                                                                                                                                                                \\ \hline
$L$                                                  & Spatial lattice                                                                                                             \\ \hline  $A$               & Set of possible states on lattice                                                                                                                          \\ \hline
$A\setminus 0$                                       & Set of epidemiological states                                                                                               \\ \hline $n_i^t$           & Total number of people at state $a_i$ at time $t$                                                                                                          \\ \hline
$\Omega_d$                                           & $d$-neighbourhood of $\mathbf{x}\in L$                                                                                      \\ \hline $\Lambda_t^a$     & Mapping $L\to a$ at time $t$                                                                                                                               \\ \hline
$p^t_{ij}$                                           & \begin{tabular}[c]{@{}l@{}}Probability at time $t$ that $\mathbf{x}\in L$ moves from $a_i$ to $a_j$\end{tabular} \\ \hline $\tau_{ij}$       & \begin{tabular}[c]{@{}l@{}}Transitional delay for  $\mathbf{x}$ to move from  $a_i$ to $a_j$\end{tabular}                                                \\ \hline
$e_t$, $i_t$                                         & \begin{tabular}[c]{@{}l@{}}Number of exposed and infected people in the\\  $d$-neighbourhood of $\mathbf{x}$ at time $t$\end{tabular}     \\ \hline $p_e$, $p_i$      & \begin{tabular}[c]{@{}l@{}}Probabilities that an exposed or an infected person spreads\\ the infection to a susceptible person when they meet\end{tabular} \\ \hline
$\Theta$                                             & A gene containing all the parameters of PCA method                                                                          \\ \hline $B$               & Binary encoded representation of $\Theta$                                                                                                                  \\ \hline
$G(\Theta)$                                          & The PCA model with parameter $\Theta$                                                                                       \\ \hline $\mathbf{y}$      & Time series of an epidemiological state in a country                                                                                                       \\ \hline
$\hat{\mathbf{y}}$ & Time series estimate of epidemiological state from PCA                                                                      \\ \hline $e_{ji}$          & Estimation error of $j^{th}$ gene in $i^{th}$ generation                                                                                                   \\ \hline
$N_g$                                                & Total number of chromosome in genepool                                                                                      \\ \hline $F$               & Number of parents selected for mating from $N_g$                                                                                                           \\ \hline
$p_\beta$                                            & Fraction of $r_t$ that recovers from the disease                                                                            \\ \hline $\rho$            & Fraction of parents F that lives in the next generation                                                                                                    \\ \hline
\end{tabular}
}}
\end{table*}
These people are denoted as removed ($R$) subpopulation in the model. This removed subpopulation contains two kinds of people$-$ one who have recovered from the infection completely and they neither infect nor get infected in future, and the other kind of people who have died due to the severity of the infection. Schematic diagram related to the transitions, probabilities and timelines corresponding to the dynamics of infection are shown in Fig. \ref{fig:scheme}(b).  In the analysis, normalized subpopulations have been considered, and the respective normalized subpopulation is denoted using the same lowercase character. For example, the normalized susceptible and infected subpopulations are denoted by $s$ and $i$ respectively. As shown in Fig. \ref{fig:scheme}(c), this epidemiological time evolution has been implemented on a 2D lattice using PCA as discussed below.
\subsection{Probabilistic Cellular Automata} \label{ca}
Let $L $ be a finite subset of $\mathbb{Z}^2$ at time instance $t$, denoted as $L\sqsubset \mathbb{Z}^2$ which defines a regular 2D lattice. Every point on this lattice $\mathbf{x}\in L$ can acquire finite number of \textit{states} $A$. In this particular problem, the set $A$ can be defined as $A=\{0,s,e,i,q,r\}$, where the terms $s$, $e$, $i$, $q$ and $r$ denote the particular possible states of infection as discussed in Sec. \ref{ssec:subpop}, and $0$ denotes no human occupant or an empty space. At time $t=0$, $n^0_i$ points are randomly selected on $L$ and assign the state $a_i$ where $i\in A$. The total initial population is defined as  $N=\sum_{i\in A\setminus 0}n^0_i$. At any instance of time $t$, $n_i^t$, $i\in A\setminus 0$ denotes the total number of the people in respective state $a_i$.\\
\noindent
For neighbourhood criteria, modified-Moore neighbourhood or $d$ -neighbourhood has been used. A finite subset $\Omega_d\sqsubset \mathbb{Z}^2$ is defined, containing the origin $\mathbf{0}=(0,0)$, and the cardinality of $\Omega_d$ is $4d(d+1)$. General probabilistic cellular automata (PCA) is a stochastic process that describes sequence of mappings $\Lambda_t^a: L\to a$, $a\in A$, where any particular state $\Lambda_t^a(\mathbf{x})$ of $\mathbf{x}\in L$ at a particular time instance $t$ is dependent on the previous states of the $d$-neighbourhood of $\mathbf{x}$, denoted as $x+\Omega_d=\{\mathbf{x}+\mathbf{\omega}: \forall \omega\in \Omega_d \}$ with certain probabilities. More precisely, in COVID-19 infection spread, $\Lambda_t^E(\mathbf{x})$ will be decided by $\Lambda_{t-1}(\mathbf{x}+\omega)$, $\forall \omega\in \Omega_d$. The other mappings $\Lambda_t^a(\mathbf{x})$, $a\in A\setminus E$, depends on the sequence of states $\Lambda_\kappa^a(\mathbf{x})$, $0\leq \kappa<t$.\\
\subsubsection{Transitional Probabilities}
\label{tranprob}
The transition probability $p_{a_ia_j}^t$  denotes the probability of transition at time $t$ from state $a_i$ to state $a_j$, where $a_i,a_j\in A$. Without any loss of generality, $p^t_{a_ia_j}$ is  denoted as $p^t_{ij}$ and transition from state $a_i$ to $a_j$ as $a_{ij}$ in the rest of the discussion for a simpler notation. In cases, where $a_i\neq a_j$, $p^t_{ij}$ is referred as state transitional probability, and if $a_i= a_j$, $p^t_{ii}$ is called as self transitional probability.\\
If a state transition $a_{ij}$,  $i\neq j$, happens in $\mathbf{x}$ at time $t$ following the transition probability $p^t_{ij}$ and the transition state $a_{ij}$ has a transitional delay $\tau_{ij}$, then
\begin{equation*}
p^t_{ij}=\left\{\begin{matrix}
0\;\;\;\;\;\;\; \text{if } t<t_{ui}+\tau_{ij}\\ 
p_{ij} \;\;\;\; \text{if } t\geq t_{ui}+\tau_{ij}
\end{matrix}\right.
\end{equation*}
where $t_{ui}$ is the time instance when transition $a_{ui}$, $u\neq i$  happened. In this infection diffusion model, only the state transitional probabilities $p^t_{se}$, $p^t_{ei}$, $p^t_{iq}$, $p^t_{qr}$ and $p^t_{ir}$ are considered to be nonzero at certain instance of time, and for all the other transitional probabilities, $\tau_{ij}$ is set to infinity, where $p_{ij}$ and $\tau_{ij}$ are user defined parameters.  However, for the transition $a_{se}$, $t_{ui}$ and $\tau_{ij}$ are set to zero, and for  $\mathbf{x}\in L$, let us define $p^t_{se}=p_{ij}=1-p^t_{ss}$ and the self-transition probability  $p^t_{ss}=(1-p_i)^{i_{t-1}}(1-p_e)^{e_{t-1}}$
where $i_{t-1}$ and $e_{t-1}$ are the number of cells in states $i$ and $e$ respectively in the $\Omega_d$ neighbourhood of $\mathbf{x}$ at time $t-1$. The probabilities $p_e$ and $p_i$ are defined as `infection probabilities' which can be considered as the probabilities that a susceptible person become exposed to the infection when that person meets an exposed or an infected person respectively. \\
An empty cell does not contribute in the infection spread, and thus, self transitional probability $p^t_{00}=1$, $\forall t$. Among the total removed population $r_t$ at time instance $t$, a population fraction $p_\beta r_t$ is considered that recover from the infection at time instance $t$ and acquire long-term immunity towards the disease, and a population fraction $(1-p_\beta) r_t$ is considered to be deceased. The removed population $r_t$ is not considered further in the infection dynamics and it is taken that $p^{t'}_{rr}=1$, $t'>t$.
\begin{figure*}[tbp]
\begin{center}
\includegraphics[width=0.879\linewidth]{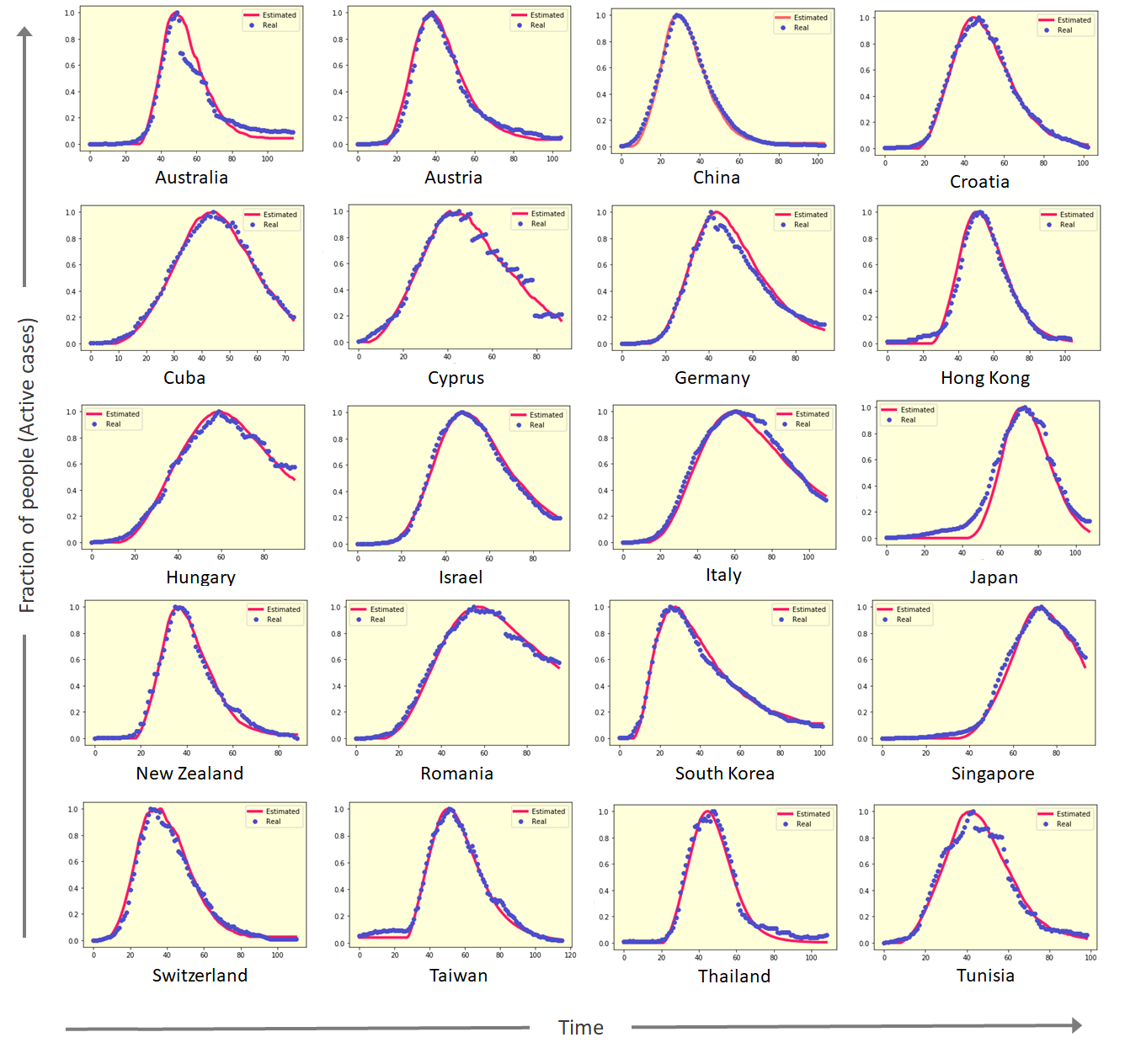}
\caption{Time series data for active cases (blue) of COVID-19 pandemic in different countries where the peaks of the infection spread of the first wave have been passed, and estimated active cases (red) from proposed PCA-GA method.}
\label{fig:active_cases}
\end{center}
\end{figure*}
\subsection{Parameter optimization using GA} \label{ssec:ga}
Though PCA has potential to model the probabilistic transition of states on a spatial lattice, the main challenge to use it for modeling a real-world scenario is to find out the optimal parameters for the PCA. As the searching space for the proposed PCA model is very large, it is practically impossible to search for the optimal parameter setting manually to analyse the characteristics of the infection spread from a real data. Thus, genetic algorithm (GA) has been applied to find out the optimal parameter set given a real time-series data.\\
Let us assume a discrete time signal $y[\text{n}]$, $0\leq \text{n}\leq (T-1)$  associated with the real world infection spread. The PCA model is denoted by $G(\Theta)$, where $\Theta=[\theta_1,\theta_2 \hdots \theta_h ]$ denotes the set of parameters used for the PCA model. If $\hat{y}[\text{n}]$, $0\leq \text{n}\leq (T-1)$ is the time evolution of the desired variable in the model $G(\Theta)$, then the objective is to find an optimal parameter set $\Theta^*$ such that $\hat{y}[\text{n}]\rightarrow y[\text{n}]$, $\forall \text{n}$. To apply GA, each $\theta_i$, $1\leq i\leq h$, is encoded as a string of binary digits $b_i$ \cite{gulsen1995genetic,karr1995least} assuming the $\theta_i$ has a bound $|\theta_i|<\zeta_i$, $1\leq i\leq h$. This binary string is referred as \textit{gene}, and the concatenated genes in the order of the appearance of respective $\theta_i$ in $\Theta$ is called the \textit{chromosome}. For example, if $B$ is the chromosome corresponding to parameter set $\Theta$, $G(B)$ is equivalent to $G(\Theta)$. A collection of $N_g$ number of chromosomes of estimated parameters, often referred as \textit{gene pool}, are evaluated at every time step (called as \textit{generation}). In our work, the error of each chromosome has been evaluated using $l_1$ norm distance. At $i^{\text{th}}$ generation, the error of the $j^{\text{th}}$ chromosome $B_{ji}$ is computed as
\begin{equation*}
    e_{ji}=\|\mathbf{y}-\hat{\mathbf{y}}_{ji}\|_1=\sum_{\text{n}=0}^{T-1}|y[\text{n}]-\hat{y}_{ji}[\text{n}]|
\end{equation*}
where $\hat{\mathbf{y}}_{ji}$ is the estimated output of $G(B_{ji})$ in the vector form and $\hat{y}_{ji}[\text{n}]$ is the value of $\hat{\mathbf{y}}_{ji}$ at time instance `$\text{n}$'. At each generation, GA finds out $min(e_{ji})$, $\forall j$ and tries to make $e_{ji}\rightarrow 0$ as $i\rightarrow \infty$. In the proposed framework, some of the parameters are related to probabilities having a range 0 to 1, and some of the parameters are associated with time (in days) which are discrete integers, and greater than or equal to zero in our case. Thus, the parameters are initialized randomly keeping their domain restrictions intact.\\
For mating, two chromosomes, often referred as \textit{parents}, are selected from the gene pool considering their `\textit{fitness}'. Among two selected parents, a crossover point or a splice point is selected at $b_i$, $1\leq i\leq h$ in both chromosomes and a crossover \cite{karr1995least} happens that produces two offsprings. In our approach, fitness $f_{ji}$ of each chromosome has been defined as the inverse of their respective errors at a particular generation. At each generation, $F$ number of best chromosomes are selected from the gene pool having the maximum fitness for mating. Following the idea of \cite{yao1994nonlinear}, $\rho F$ number of parents are kept to the next generation along with the new chromosomes to ensure that the error in the next generation is always less than or equal to the current generation. Selecting $\rho F$ number of chromosomes from the parents, $N_g-\rho F$ number of children are produced from mating to keep the size of the gene pool constant. After the offsprings are generated, in the parameter space, $s$ genes are randomly selected and small perturbations are added individually to mimic mutation.\\
As shown by several researchers \cite{holland1992adaptation}, the homogeneity in the gene pool increases with the generations, and as the perturbations due to mutation are typically small, the reduction of error becomes a problem after a few generations. Thus, to restrict homogeneity in the gene pool, a small number of offsprings $\mu$ are selected from the total  $N_g-\rho F$ number of generated offsprings, and replaced them with randomly generated chromosomes to maintain diversity. This step is called as `diversification' of gene pool.\\
\begin{figure}
\begin{center}
\includegraphics[width=\linewidth]{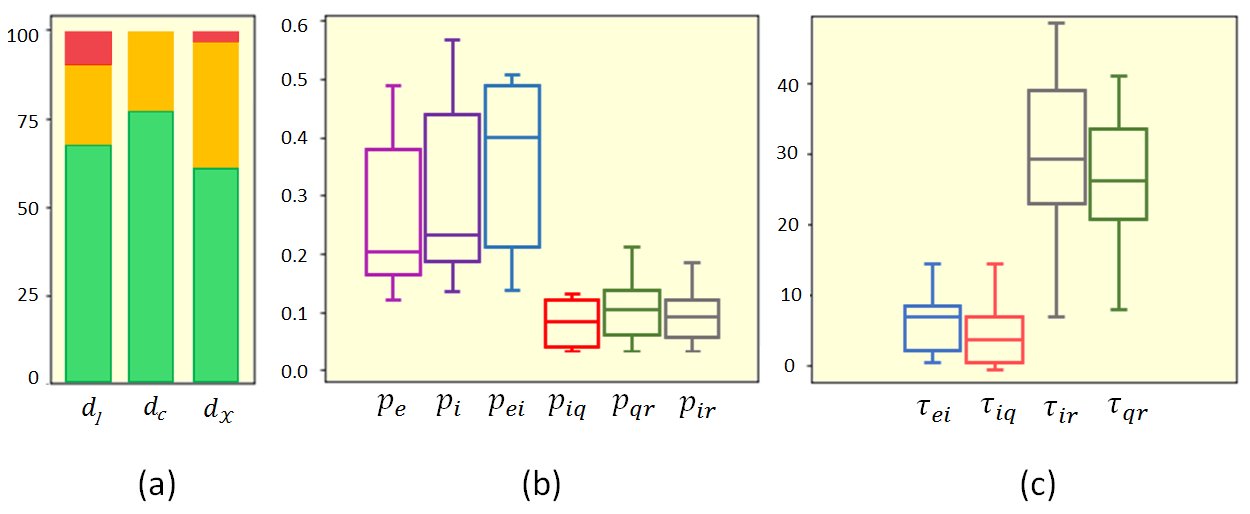}
\caption{Parameter estimations and goodness of model estimation: (a) RMSE, Correlation and $\chi^{2}$ distance,  $d_l$, $d_c$ and $d_{\chi}$ for all 40 countries considered in this work in terms of goodness of agreement with model estimations shown in percentage. The colors green, orange and red signify level of agreement. Values between (0:0.05) for $d_l$, (0:0.01) for $d_c$ and (0:1) for $d_\chi$ are considered as good (green). Values between (0.05:0.08) for $d_l$, (0.01:0.1) for $d_c$ and (1:3) for $d_\chi$ are considered as moderate (orange). Values above moderate are considered as poor (red). For all three metrics $65-75\%$ countries have shown good agreement with model estimation; (b) and (c) represent boxplot for the best-fit parameters of state transition probabilities and state transitional delays respectively, for all the 20 countries shown in Fig. \ref{fig:active_cases}. The height of the boxplots represents the interquartile range (IQR). The dark line inside the box represents the median. The lower and upper whisker extend to the lowest and highest values within 1.5 IQR of the first and third quartile, respectively.}
\label{fig:stat}
\end{center}
\end{figure}
In our problem, the parameters $\Theta$ of the PCA model $G(\Theta)$ are the state transitional probabilities $p_{ei}$, $p_{iq}$, $p_{ir}$, $p_{qr}$, infection probabilities $p_e$ and $p_i$, state transition delays $\tau_{ei}$, $\tau_{iq}$, $\tau_{qr}$, $\tau_{ir}$, neighbourhood $d$, and death probability $p_\beta$ as mentioned in Sec.\ref{ca}. As optimizing these many parameters simultaneously might be challenging and require huge amount of resources, we propose a variant of GA with sequential evolution mechanism where instead of optimizing the solutions simultaneously,  the parameters are  optimized sequentially. Let us define a set of generations as an \textit{era}. For the first era containing a small number of generations, a traditional GA methodology is followed as discussed this far to have a set of initial parameters. From the next era onward, two parameters are fixed and optimized sequentially in that era. Mutation and crossover are restricted to those two respective genes, whereas parent selection is done based on the performances of the entire chromosomes. This newly proposed sequential optimization of parameters of PCA using GA is defined as PCA-GA. The proposed approach can optimize a large number of parameters using limited resources efficiently. All the notations used in PCA-GA are briefly summarized in the Table \ref{params}.\\
{Proposed PCA-GA has a complexity which can be  approximated as $O(N_gT_gO(\mathit{f}))$ where $N_g$ is the number of population, $T_g$ is the total generation and $O(\mathit{f})$ is the complexity to measure the fitness in the GA. For a large enough $N_g$, $T_g$ is considered as a comparatively smaller constant and thus, the complexity of the entire algorithm is mainly governed by $N_g$ and $O(\mathit{f})$. The complexity of estimating the fitness can be approximated as $O(\mathit{f})=O(T+8N\tau T)$ for Moore neighbourhood criteria, where $N$ is the total population on the 2D grid.  The length of the original time series data $T$, and $\tau$, the maximum of $\tau_{ij}$, are both constant, and thus $O(\mathit{f})$ can be represented as $O(N)$.}\\
{Though GA has been selected as a strategy to optimize the parameters of the proposed PCA model, it is evident that because of the generalized construction of the proposed framework, other meta-heuristic methods could also be employed to search the parameters of the spatially driven SEIQR model which is the main focus of this work. However, presence of mutation and diversification in GA help to search for better solutions as the search space is extremely large.}
\begin{figure}
\begin{center}
\includegraphics[width=\linewidth]{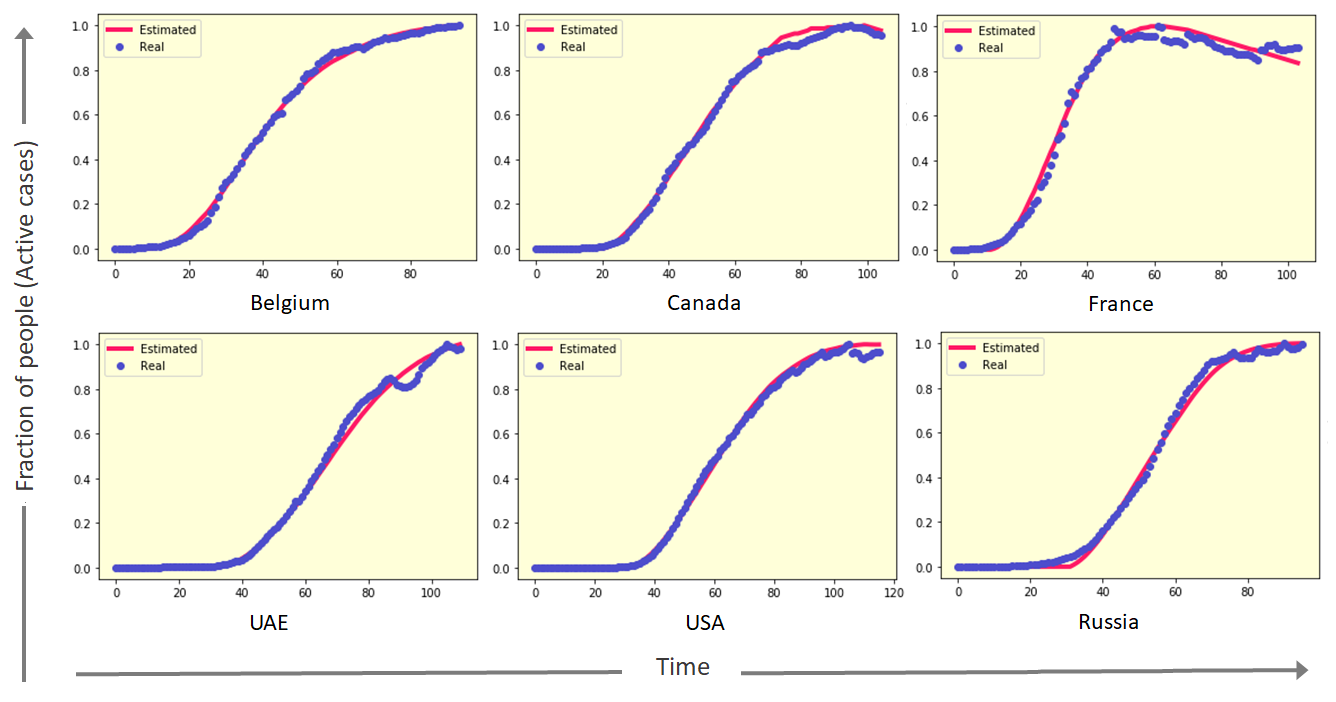}
\caption{Time series data for active cases (blue) of COVID-19 pandemic in different countries where the cases are saturating, and estimated active cases (red) from proposed PCA-GA method.}
\label{fig:active_cases2}
\end{center}
\end{figure}
\section{Results}
To validate the effectiveness of the proposed framework, using PCA-GA, the actual statistics of COVID-19 spreads till 20th June, 2020 in different countries is used. {For finalizing the data-set from available data of 213 countries, several aspects have been considered. At first, 102 countries had been dropped due to less number of reported cases (less than 1000 reported cases till 20th June 2020). Out of the remaining countries, some countries, like Iran, Greece, Paraguay etc., are removed due to data inconsistency, and finally 40 countries are randomly selected ensuring the following points:} 
\begin{itemize}
    \item At least 2 countries from each continent got selected to maintain demographic diversity in our data.
    \item Care has been taken to maintain significant variation in population density, which we believe as a major factor contributing in disease transmission.
    \item It was ensured that countries from three distinct stages of COVID-19 infection are considered: (i) where the infection is significantly diminished, (ii) where the peak infection has been reached but substantial infection still persists, and (iii) where consistent growth in infection is occurring.
\end{itemize}   
With these widely variant spectrum of time series data, we proceed for quantitative calibration and interpretation through the proposed methodology. All data samples are taken from the website worldometers.info\footnote{\url{https://www.worldometers.info/coronavirus/}}.  \\
 To point out the major contributing factors in dynamics of infection spread, for every country under consideration, three available time series, namely daily active cases, total number of infected cases and total number of deaths are accumulated. Out of these three series, the daily active cases time series is used for model formulation, and the rest are considered for model validation. It is important to mention that the population $q_t$ is the relevant observable here, as infected people as $i_t$ and $e_t$ remain latent and undetected in the population. The reported daily active case data is associated with lifetime of the infection, and are used in this study to check the effectiveness of the proposed framework as follows. By applying PCA-GA on the daily active case data of a particular country, the parameters $\Theta^*$ that gives the minimum $l_1$ error is extracted. {For validation of the optimized parameters and understanding the robustness of the algorithm, results generated by using $G(\Theta^*)$ for the total infected states and deceased states are then compared with the real-world data. Here it must be noted that the optimal parameters $\Theta^*$ remain unaltered and no further optimization is performed.}
\subsection{Experimental Setup}
For all the simulations, PCA is initialized with a fixed lattice size of $100\times 100$ with $n_e=50$ and $n_i=4$. The population $n_q$ and $n_r$ are set to zero at $t=0$. The susceptible population $n_s$ has been initiated depending on the population density of a country as follows: among the countries considered in our study, for the country with lowest population density (Canada), $n_s=2500$ has been selected, and for the country with highest population density (Singapore), $n_s=6000$ has been fixed. For any other country, $n_s$ has been assigned within this range using logarithmic scaling based on the population of that country. As each of the parameters of PCA-GA has physical relevance, the sequential searching process has been initiated by following restrictions of ranges. It is important to note that in our problem, genes associated with probabilities are initiated in the range [$0,1$] and clipped during the optimization process accordingly. The state transition delays $\tau_{ei}$ (incubation period) and $\tau_{iq}$ (testing delay) are considered to be within the range ($0,30$). The transition delay $\tau_{ir}$ and $\tau_{qr}$ (corresponding recovery periods) are initialized in the range ($20,100$). {All the simulations are executed in a system with Intel Core i7 8700K processor, 64GB RAM and 8GB NVIDIA GeForce RTX 2080 8GB GPU using Python and numpy packages.}
\begin{figure}
\begin{center}
\includegraphics[width=\linewidth]{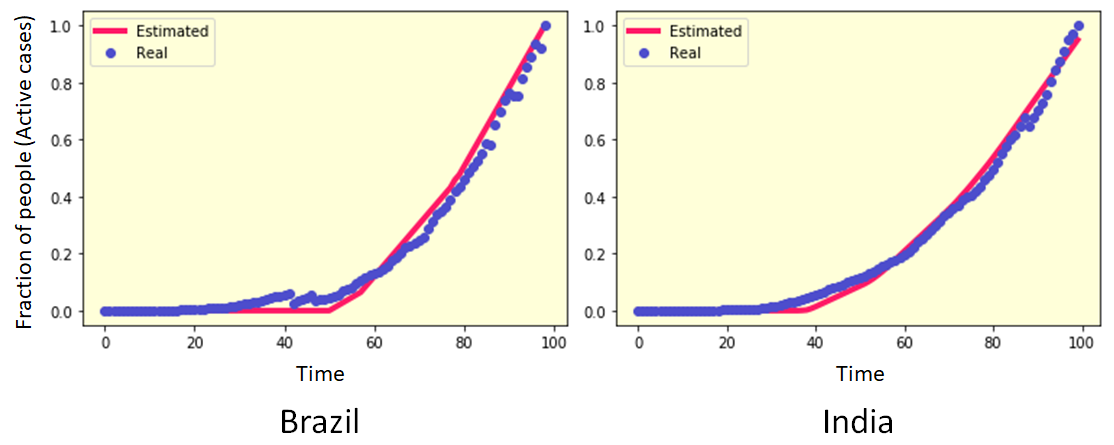}
\caption{Time series data for active cases (blue) of COVID-19 pandemic in different countries where the cases are increasing exponentially, and estimated active cases (red) from proposed PCA-GA method.}
\label{fig:active_cases3}
\end{center}
\end{figure}
\subsection{Estimation of Parameters Using Active Cases:}\label{ssec:active}
The daily active cases can be defined as the $c_t=c_{t-1}+q_t-r_t$ where $c_t$ is the number of active cases at time instance $t$ having the initialization $c_0=0$. In Fig. \ref{fig:active_cases}, the active cases of 20 different countries are shown along with the respective estimated active cases using PCA-GA model. For the countries shown in Fig. \ref{fig:active_cases}, the first peak of the infection is already crossed and a steady fall in the infection spread is observed. It can also be seen that some of the active cases of the countries like China, Israel, Switzerland, follow smooth bell-shaped curves, whereas for some countries, like Australia, Cyprus, Hungary etc., the times series data deviates from bell-shaped curves with substantial degree of noises. In all the cases, PCA-GA has successfully captures the trend of the time series data estimating the parameters of the epidemiological process. {To measure the goodness of the model estimation, three different metrics has been used to measure the quality of the estimated values. The root mean square (RMSE) distance, correlation distance and chi-square distance \cite{liao2005clustering,gao2009denoising,salem2014anomaly}, denoted as $d_l$, $d_c$ and $d_{\chi}$ respectively, are computed between the real data and the estimated values from the PCA-GA model to evaluate the effectiveness of the optimized model. For two vectors $\mathbf{u}$ and $\mathbf{v}$, we define
\begin{eqnarray}
d_l=\sqrt{\frac{1}{T}\sum_{i=1}^T{(u_i-v_i)}^2},\; 
d_c= 1-\frac{(u-\bar{u}).(v-\bar{v})}{\|(u-\bar{u})\|_2\|(v-\bar{v})\|_2},\;
d_{\chi}=\sum_{i=0}^T\frac{(u_i-v_i)^2}{v_i} \nonumber
\end{eqnarray}
where $T$ is the length of each vector, $u_i$ and $v_i$ are the $i^{th}$ elements of $\mathbf{u}$ and $\mathbf{v}$ respectively and (.) denotes dot product of two vectors.
As shown in Fig. \ref{fig:stat}(a), the proposed model performs well in modelling the real data. When evaluated over all the countries considered in this work, the proposed model fits the data well, and for only 0\% -12.5\% cases the fittings were poor depending on the evaluation metric.  It is important to mention that all the distance measures are evaluated on normalized data.}\\ 
In Fig.  \ref{fig:active_cases}, an interesting point to notice is that the peak of the active cases are located at markedly differing time instances, and the other properties, like variance, skewness etc., of the observed distributions are also varying drastically. The fundamental differences between the fitted curves are quantified with the help of boxplot of the parameters in Fig.\ref{fig:stat}(b)-(c) by analysing basic statistical properties. The reported boxplots are specifically for the countries selected in Fig.  \ref{fig:active_cases}. It can be noted that $p_{e}$, $p_{i}$ and $p_{ei}$ exhibit a wide variability in Fig. \ref{fig:stat}(b). During our analysis, a strong positive correlation with population density for $p_{e}$ and $p_{i}$ has been also observed. This can be thus inferred that the variation in population density in the considered countries causes the wide range of these parameters. It can be also concluded that high density of population increases the probability of transmission of the disease. The considerable difference in the mean magnitudes of the infection associated probabilities ($p_{e}$, $p_i$ and $p_{ei}$) and recovery-related probabilities ($p_{iq}$, $p_{ir}$ and $p_{qr}$) indicate the sharper rise and slower fall of active cases curves, which results into a skewed distribution in most of the cases (see Fig. \ref{fig:active_cases}). In Fig. \ref{fig:stat}(c), it is also shown that $\tau_{ei}$, which is identified as the incubation time in the model, exhibits a range of 3-14 days with a mean at 7.3, which perfectly aligns with the observed cases all around the world \cite{who2020incubation}. In this figure, a wide variability in the range of $\tau_{ir}$ and $\tau_{qr}$ is observed, which points out the substantial difference in health infrastructure of these countries.\\
\begin{figure}
\begin{center}
\includegraphics[width=\linewidth]{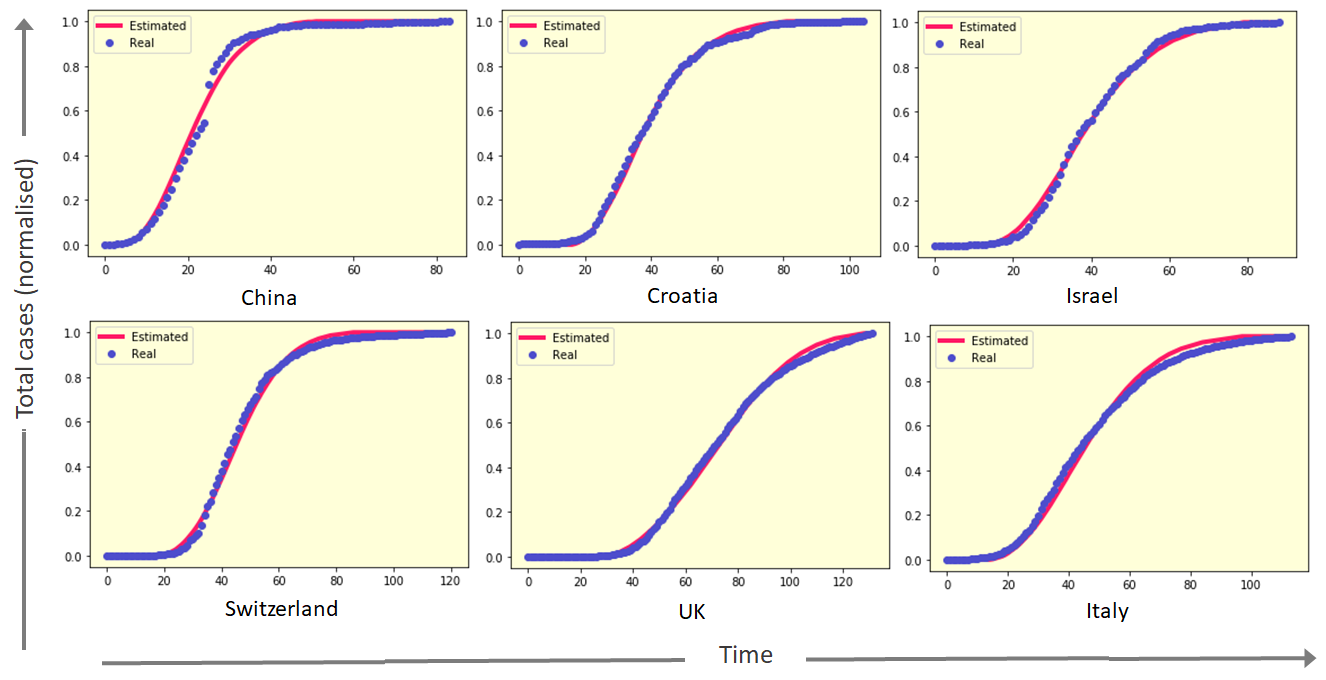}
\caption{Total infected cases (blue) of COVID-19 pandemic in different countries, and estimated total cases (red) from proposed PCA-GA method.}
\label{fig:total_cases}
\end{center}
\end{figure}
Here it must be mentioned that, while performing this statistical analysis with all 40 countries, some countries were detected showing consistent outliers (not included in Fig. \ref{fig:stat}(b)-(c)) in terms of four transitional parameters: $p_{ir}$. $p_{qr}$, $\tau_{ir}$ and $\tau_{qr}$. While analyzing the active case distributions of these outliers, it was found out that the time series data for all these countries have a saturating trend where the daily active cases do not show an average descent with time. Some of such cases are shown in Fig. \ref{fig:active_cases2}. Even for these data which have drastically different qualitative trend compared to countries shown in Fig.  \ref{fig:active_cases}, the proposed PCA-GA framework has successfully captured the trend of the real time series data accurately.\\
There are also certain countries, like India, Brazil, Chile, Mexico, etc., for which the infection spreading started later than the countries like China or Italy, and the active daily cases are still growing almost exponentially. As shown in Fig. \ref{fig:active_cases3}, PCA-GA is able to estimate the time series data for these countries where the infection is spreading rapidly. Dynamics of COVID-19 spread in these countries are of particular interest as the prediction of the peak positions in these countries might help immensely to understand the maximum socioeconomic impact of the disease at a time in that geographical location. 
\begin{figure}
\begin{center}
\includegraphics[width=\linewidth]{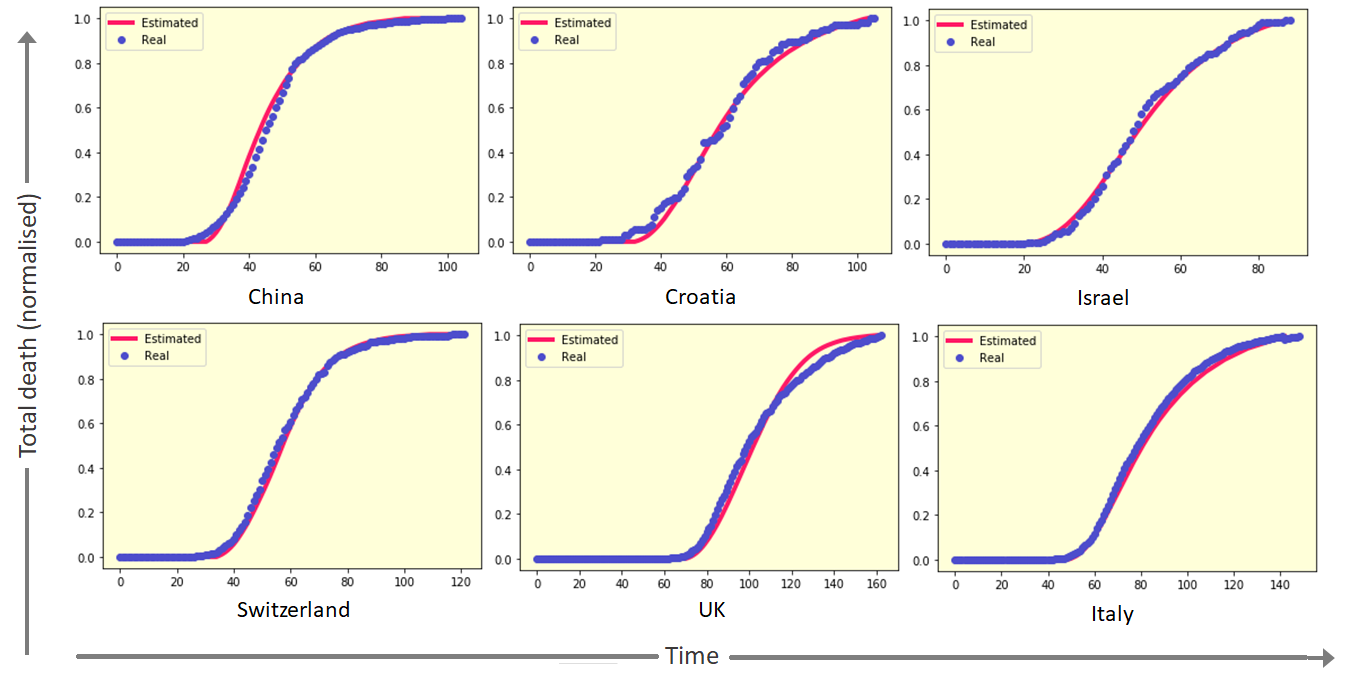}
\caption{Total deaths (blue) of COVID-19 pandemic in different countries, and estimated total deaths (red) from proposed PCA-GA method.}
\label{fig:death_cases}
\end{center}
\end{figure}
\subsection{Validation of the Proposed Model}
{While analyzing a complex dynamics like the spread of a pandemic, it is not always sufficient to model the input real data only. It is required that the optimized model should be robust and can provide meaningful interpretations without further retraining or parameter tuning for real-world applications. To validate the robustness and the effectiveness of the proposed algorithm, the optimized model is now employed for three different tasks. At first, the robustness of the optimized model is checked by estimating the total number of infected cases, followed by total number of death cases without any further training, tuning or supervision. Finally, to further validate the efficiency of the model, its performance has been evaluated for the prediction task by training the model with partitioned data and evaluating on its future predictions without any further optimization.} 
\subsubsection{Total Number of Infected:}
The total number of infected cases $z_t$ at time instance `$t$' can defined as $z_t=\sum_{i=0}^t q_i$. This cumulative sum indicates the total number of people who suffered from the disease at any point of time. For a country, where the first wave of the infection has passed, e.g., Croatia, Italy, etc., $z_t$ follows a sigmoid function approximately, whereas for the countries like India, Mexico etc., where the infection has not reached the peak, $z_t$ follows an exponential function. As PCA-GA is optimized using the time series information of daily active cases $c_t$, $z_t$ is used to validate the parameters learnt by the sequential GA framework in the following way. Once a particular country is selected, $\Theta^*$ is estimated using PCA-GA with the actual $c_t$. Next the $\hat{z}_t$ for $G(\Theta^*)$ is calculated without any further fine-tuning of the parameters, and compared $\hat{z}_t$ with actual $z_t$. In Fig. \ref{fig:total_cases}, the total cases (blue) of six such countries are shown along with the best-fit results obtained from PCA-GA (red) which depict an excellent agreement with the data. It must be mentioned that for all three dynamical stages of infection spreading as discussed in Sec. \ref{ssec:active}, i.e., where the first wave of infection has passed, where the active cases are almost saturated currently or where the active cases are increasing rapidly, our estimated $\hat{z}_t$ closely matches $z_t$ without any further parameter optimization. {When evaluated over all 40 countries for the number of infected people, the proposed method gives average $d_l$, average $d_c$ and average $d_\chi$ as 0.037,0.006 and 0.53 respectively, which exhibits the robustness of the model.}
\subsubsection{Total Death Cases:} 
To further validate the `goodness' of the estimated parameters, the parameter set $\Theta^*$ optimized over the daily active cases of a particular country is taken and the identical parameter values are used to compare the estimated total deaths with the actual total deaths of that country. Death in the population is the prime concern in case of the COVID-19 pandemic, and as mentioned in Sec. \ref{tranprob}, daily deceased population is a fraction of $r_t$ in our model. So, the total estimated death cases can be defined as $\hat{d}_t=(1-p_\beta)\sum_{i=0}^t r_i$ where $p_\beta$ and $r_i$ for $0\leq i\leq t$ are given by $\Theta^*$ and $G(\Theta^*)$ respectively. Fig. \ref{fig:death_cases} demonstrates the comparison of the actual total death cases $d_t$ with estimated total death cases  $\hat{d}_t$ for $\Theta^*$, the identical set of parameters used for estimating active cases as well as total cases previously. The same countries shown in Fig. \ref{fig:total_cases} have been selected to show the robustness of the estimated parameter $\Theta^*$ using the proposed technique. {Excellent agreement with data has been found for this case as well; when evaluated over all 40 countries for the total number of death cases, the proposed method gives average $d_l$, average $d_c$ and average $d_\chi$ as 0.041,0.006 and 0.48 respectively}.
\subsubsection{Prediction Related to Infection Spread:}
Prediction of future events is always challenging in data modeling \cite{acharjya2017comparative}.For the final stage of validation of the methodology, the predictive power of the model has been tested. As the impacts of this pandemic becomes far reaching as the socioeconomic contexts vary, a considerably accurate prediction about the dynamics of the infection spread can be crucial and useful in many ways. As PCA-GA successfully estimates the optimal parameter $\Theta^*$, the set of parameters can also be utilised to predict the future course of the infection in that country. 
\begin{figure}
\begin{center}
\includegraphics[width=\linewidth]{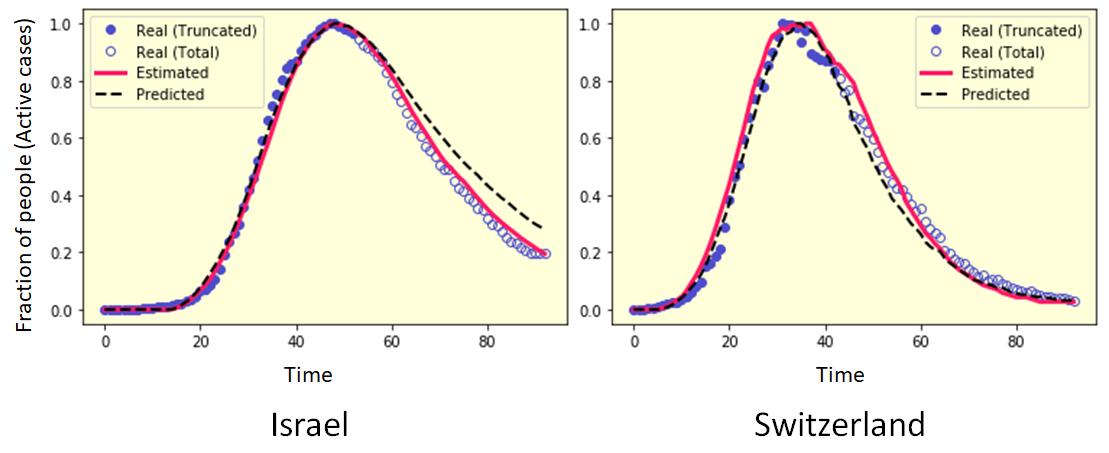}
\caption{Prediction of daily active cases from truncated data. For Israel and Switzerland, real data up to 54 and 43 days has been used to predict the daily active cases for 100 days. For prediction, the average of 50 independent PCA-GA simulations are considered.}
\label{fig:pred_val}
\end{center}
\end{figure}
To validate the capacity of the prediction strategy, the daily active cases of a country $c_t$ is truncated to $c_P$ keeping the first `$P$' values. PCA-GA is applied on $c_P$ to estimate the parameters $\Theta_P$. Then $\Theta_P$ is used to predict the daily active cases $\hat{c}_t$. As shown in Fig. \ref{fig:pred_val}, for two countries Israel and Switzerland, the daily active case information up to 54 and 43 days respectively are considered for an attempt to predict the daily active cases up to 100 days. In the figure, the estimated curve (shown in red) is optimized using all the real data points available, whereas the predicted curve (shown in black) is optimized using the truncated real data. It can be observed that the predictive estimation closely follows the real active case data, even though only $\sim 50\%$ data points are used for parameter estimation. For Israel and Switzerland, 100 days prediction of the algorithm produces $(d_l,d_c,d_\chi)$ as $(0.056,0.008,0.95)$ and $(0.028,0.005,0.43)$ respectively. {As prediction of the spread of the infection is one of the most challenging tasks, the predictive ability of proposed algorithm is compared with different baseline methods to better understands its performance. As only a very few data points were available in the truncated data, fast decision tree learning algorithm \cite{su2006fast} and Random forest regression perform poorly and give $(d_l,d_c,d_\chi)$ as $(0.43,0.49,243.82)$ and  $(0.439,0.51,252.6)$ respectively for the truncated time series of Switzerland. SVM regression with RBF kernel performs satisfactorily on the same truncated data and produces $(d_l,d_c,d_\chi)$ as $(0.09,0.02,27.8)$. However, the proposed PCA-GA algorithm significantly outperforms the baseline algorithms and produces  $(d_l,d_c,d_\chi)$ as $(0.028,0.005,0.43)$.}
\subsection{Prediction for exponentially rising active cases}
As the PCA-GA methodology has been elaborately validated in Section 3.3, now, in this section, it is employed for the purpose of prediction of consistently rising real epidemic data. Though the parameter estimation works well
even when the minimum information about the peak position in $c_t$ is available, the prediction task becomes really challenging when $c_t$ is exponential in nature. For a particular country where $c_t$ is almost exponentially rising, proceeding with prediction, first the best set of parameters $\Theta^*$ is detected by PCA-GA with fitness $f^*$ and error $e^*$. As the drop of the infection heavily depends on the transitional probabilities $p_{ir}$, $p_{qr}$ and state transitional delays $\tau_{ir}$ and $\tau_{qr}$, this parameters are tuned to find a region of predictions bounded by the possible best case and the worst case scenarios. While estimating the best case scenario, $p_{ir}$ and $p_{qr}$ is chosen equal to the maximum and minimum $p_{ir}$ and $p_{qr}$ observed in the continent from which the country belongs. The reason behind this strategy is that the parameters related to the infection spreading are different in each continent which is also observed by \cite{miller2020correlation}. In the best case scenario, transitional delays $\tau^*_{ir}$ and $\tau^*_{qr}$ are reduced to obtain best case transitional delays $\tau^\ominus_{ir}$ and $\tau^\ominus_{qr}$ respectively such that the fitness remain within $90\%$ of $f^*$, where $\tau^*_{ir}$ and $\tau^*_{qr}$ are the corresponding optimized delays available in $\Theta^*$. For the worst case scenario, we consider $\tau^\oplus_{ir}=\tau^*_{ir}+\alpha_{ir}$ and $\tau^\oplus_{qr}=\tau^*_{qr}+\alpha_{qr}$, where $\alpha_{ir}=\tau^*_{ir}-\tau^\ominus_{ir}$ and $\alpha_{qr}=\tau^*_{qr}-\tau^\ominus_{qr}$.\\
\begin{figure}
\begin{center}
\includegraphics[width=0.7\linewidth]{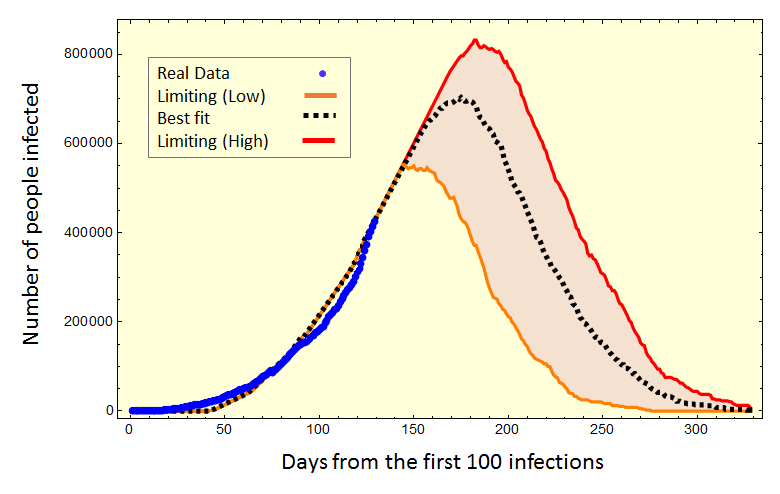}
\caption{Prediction of the course of the disease: Exponentially rising daily active cases for India (blue) till 20th July, 2020 are used for parameters estimation and the predictions.}
\label{fig:pred}
\end{center}
\end{figure}
Fig. \ref{fig:pred} depicts the prediction of the daily active cases using the method discussed so far. In the Fig. \ref{fig:pred}, the black dotted line indicates the prediction using the optimal parameters $\Theta^*$ estimated using PCA-GA. The orange line indicates the best case scenario, where the maximum daily active cases would be minimized given the real data. The red line indicates the worst case scenario based on the specific conditions mentioned above. The best case and the worst case scenarios act as limiting cases of an area (shaded in pink color) of probable future state. Any curve inside the pink region that contains the real data could be the  evolution of the daily active cases in future given the real time series data, that is in exponentially rising state currently. This indicates that for India, which is now one of the biggest epicenters of COVID-19 in South-eastern Asia, the disease can start decline very soon if vigorous measures from government and complete support from the public could be achieved. It also shows that the maximum active cases on a day, that puts a direct burden on the health infrastructure of the country can be restricted below 750,000 if people participate to government indicated mitigation strategies, and recovery rate remains at its current value. In that case, the peak of the disease is expected to pass during mid-September to mid-October, and the disease can be over with its first wave by March 2021. But these predictions also imply that the range of future states, that are possible for exponentially rising daily active cases, not only depend on the evolution of the epidemic so far, but also gets highly affected by the consistency and implementation efficiency of mitigation strategies. 
\section{Conclusion}
COVID-19 outbreak has created a massive impact all across the globe. Even after nation-wide lockdowns, extensive testing strategies and medical supports, the spread of the virus has overwhelmed several countries. Thus, it is becoming more and more important to understand the nature of the infection spread and the key parameters that are controlling the spread. In this work, we proposed a probabilistic cellular automata model to understand and depict COVID-19 spread using appropriate choice of loss functions and evolutionary optimization framework. The parameters of this cellular automata model are optimised using sequential evolutionary genetic algorithm. It has been shown that this self-adapting methodology can be highly flexible and has the power to accurately estimate time trajectories of epidemics. This model works with physically interpretable parameters, which are accessible for analysis, data collection and further experiment, and can be readily identified with ground reality. This model has been successfully employed for optimizing all these parameters simultaneously for the daily active cases, total infected cases and total deaths with extreme robustness. The performance of the model has been exhibited for a large number of countries with huge diversity in  population density, continents and available healthcare infrastructures. The predictive strength of the model has also been validated extensively, and demonstrated to estimate the course of the pandemic for the countries where infection peak has not been reached yet. It is important to mention that the motivation of the work was to develop a data driven, generalized, spatial framework that can be used to estimate relevant epidemiological parameters. This methodology is so powerful and flexible  that physical interpretations of the results obtained from these analyses can have a wide range implications. Once the data is properly interpreted with the proposed methodology, interesting realistic features can be identified for specific countries. For example, in a pandemic situation, easily relatable factors like population clusters, variable population density, variable health facilities at different places of a country etc, can be studied to understand and predict emergence of new hotspots which can be used to design selective area containment strategies. While we propose and establish the applicability and strength of this framework in this work, we wish address these application perspectives in a study in our upcoming research studies.\\
 With this proposed platform, the impact of individuality on contagion process can be explicitly studied, which might be directly related to the questions like lockdown behavioral differences, influence of rumors, vaccination opinion differences etc. {As the effects of more complex dynamical factors like periodic lockdown or population clusters are not considered in this present model, the prediction capability of the proposed model is not satisfactory for time series data with abrupt discontinuities in the present form. The proposed framework could be enhanced with other $l_p$ norm distances and different optimization techniques like multi-objective genetic algorithm or strength pareto evolutionary algorithm. Other swarm-based optimization techniques can also be explored for further refinement of the model}. The potential of the proposed approach can be utilized to better understand the disease spreading and controlling, beyond this pandemic the world is facing currently, by keeping  track of the spatial information of the dynamics, incorporating realistic behavioural aspects, and optimizing in terms of demographic as well as socioeconomic features.
\bibliographystyle{elsstyle}      
\bibliography{bibliography}  

\begin{thebibliography}{64}
\expandafter\ifx\csname natexlab\endcsname\relax\def\natexlab#1{#1}\fi
\providecommand{\bibinfo}[2]{#2}
\ifx\xfnm\relax \def\xfnm[#1]{\unskip,\space#1}\fi
\bibitem[{{\relax{World Health Organization Coronavirus disease (COVID-2019)
  situation reports}}(2020)}]{who2020}
\bibinfo{author}{{\relax{World Health Organization Coronavirus disease
  (COVID-2019) situation reports}}}, \bibinfo{title}{Available at url:
  \url{https://www.who.int/emergencies/diseases/novel-coronavirus-2019/situation-reports}
  (accessed {J}une 2020)}, \bibinfo{year}{2020}.
\bibitem[{Jin et~al.(2020)Jin, Lian, Hu, Gao, Zheng, Zhang, Hao, Jia, Cai,
  Zhang et~al.}]{jin2020epidemiological}
\bibinfo{author}{X.~Jin}, \bibinfo{author}{J.-S. Lian}, \bibinfo{author}{J.-H.
  Hu}, \bibinfo{author}{J.~Gao}, \bibinfo{author}{L.~Zheng},
  \bibinfo{author}{Y.-M. Zhang}, \bibinfo{author}{S.-R. Hao},
  \bibinfo{author}{H.-Y. Jia}, \bibinfo{author}{H.~Cai}, \bibinfo{author}{X.-L.
  Zhang}, et~al.,
\newblock \bibinfo{title}{Epidemiological, clinical and virological
  characteristics of 74 cases of coronavirus-infected disease 2019 ({COVID}-19)
  with gastrointestinal symptoms},
\newblock \bibinfo{journal}{Gut} \bibinfo{volume}{69} (\bibinfo{year}{2020})
  \bibinfo{pages}{1002--1009}.
\bibitem[{Pan et~al.(2020)Pan, Mu, Yang, Sun, Wang, Yan, Li, Hu, Wang, Hu
  et~al.}]{pan2020clinical}
\bibinfo{author}{L.~Pan}, \bibinfo{author}{M.~Mu}, \bibinfo{author}{P.~Yang},
  \bibinfo{author}{Y.~Sun}, \bibinfo{author}{R.~Wang},
  \bibinfo{author}{J.~Yan}, \bibinfo{author}{P.~Li}, \bibinfo{author}{B.~Hu},
  \bibinfo{author}{J.~Wang}, \bibinfo{author}{C.~Hu}, et~al.,
\newblock \bibinfo{title}{Clinical characteristics of {COVID}-19 patients with
  digestive symptoms in {H}ubei, {C}hina: a descriptive, cross-sectional,
  multicenter study},
\newblock \bibinfo{journal}{The American journal of gastroenterology}
  \bibinfo{volume}{115} (\bibinfo{year}{2020}).
\bibitem[{Cheng et~al.(2020)Cheng, Luo, Wang, Zhang, Wang, Dong, Li, Yao, Ge,
  and Xu}]{cheng2020kidney}
\bibinfo{author}{Y.~Cheng}, \bibinfo{author}{R.~Luo},
  \bibinfo{author}{K.~Wang}, \bibinfo{author}{M.~Zhang},
  \bibinfo{author}{Z.~Wang}, \bibinfo{author}{L.~Dong},
  \bibinfo{author}{J.~Li}, \bibinfo{author}{Y.~Yao}, \bibinfo{author}{S.~Ge},
  \bibinfo{author}{G.~Xu},
\newblock \bibinfo{title}{Kidney disease is associated with in-hospital death
  of patients with {COVID}-19},
\newblock \bibinfo{journal}{Kidney International}  (\bibinfo{year}{2020}).
\bibitem[{Han et~al.(2020)Han, Duan, Zhang, Spiegel, Shi, Wang, Zhang, Lin,
  Liu, Ding et~al.}]{han2020digestive}
\bibinfo{author}{C.~Han}, \bibinfo{author}{C.~Duan},
  \bibinfo{author}{S.~Zhang}, \bibinfo{author}{B.~Spiegel},
  \bibinfo{author}{H.~Shi}, \bibinfo{author}{W.~Wang},
  \bibinfo{author}{L.~Zhang}, \bibinfo{author}{R.~Lin},
  \bibinfo{author}{J.~Liu}, \bibinfo{author}{Z.~Ding}, et~al.,
\newblock \bibinfo{title}{Digestive symptoms in {COVID}-19 patients with mild
  disease severity: clinical presentation, stool viral {RNA} testing, and
  outcomes},
\newblock \bibinfo{journal}{The American journal of gastroenterology}
  (\bibinfo{year}{2020}).
\bibitem[{Zheng et~al.(2020)Zheng, Ma, Zhang, and Xie}]{zheng2020covid}
\bibinfo{author}{Y.-Y. Zheng}, \bibinfo{author}{Y.-T. Ma},
  \bibinfo{author}{J.-Y. Zhang}, \bibinfo{author}{X.~Xie},
\newblock \bibinfo{title}{{COVID-19} and the cardiovascular system},
\newblock \bibinfo{journal}{Nature Reviews Cardiology} \bibinfo{volume}{17}
  (\bibinfo{year}{2020}) \bibinfo{pages}{259--260}.
\bibitem[{Wang et~al.(2020{\natexlab{a}})Wang, Pan, Wan, Tan, Xu, Ho, and
  Ho}]{wang2020immediate}
\bibinfo{author}{C.~Wang}, \bibinfo{author}{R.~Pan}, \bibinfo{author}{X.~Wan},
  \bibinfo{author}{Y.~Tan}, \bibinfo{author}{L.~Xu}, \bibinfo{author}{C.~S.
  Ho}, \bibinfo{author}{R.~C. Ho},
\newblock \bibinfo{title}{Immediate psychological responses and associated
  factors during the initial stage of the 2019 coronavirus disease ({COVID}-19)
  epidemic among the general population in {C}hina},
\newblock \bibinfo{journal}{International journal of environmental research and
  public health} \bibinfo{volume}{17} (\bibinfo{year}{2020}{\natexlab{a}})
  \bibinfo{pages}{1729}.
\bibitem[{Wang et~al.(2020{\natexlab{b}})Wang, Wang, Chen, and
  Qin}]{wang2020unique}
\bibinfo{author}{Y.~Wang}, \bibinfo{author}{Y.~Wang},
  \bibinfo{author}{Y.~Chen}, \bibinfo{author}{Q.~Qin},
\newblock \bibinfo{title}{Unique epidemiological and clinical features of the
  emerging 2019 novel coronavirus pneumonia (covid-19) implicate special
  control measures},
\newblock \bibinfo{journal}{Journal of medical virology} \bibinfo{volume}{92}
  (\bibinfo{year}{2020}{\natexlab{b}}) \bibinfo{pages}{568--576}.
\bibitem[{van Doremalen et~al.(2020)van Doremalen, Bushmaker, Morris, Holbrook,
  Gamble, Williamson, Tamin, Harcourt, Thornburg, Gerber
  et~al.}]{van2020aerosol}
\bibinfo{author}{N.~van Doremalen}, \bibinfo{author}{T.~Bushmaker},
  \bibinfo{author}{D.~H. Morris}, \bibinfo{author}{M.~G. Holbrook},
  \bibinfo{author}{A.~Gamble}, \bibinfo{author}{B.~N. Williamson},
  \bibinfo{author}{A.~Tamin}, \bibinfo{author}{J.~L. Harcourt},
  \bibinfo{author}{N.~J. Thornburg}, \bibinfo{author}{S.~I. Gerber}, et~al.,
\newblock \bibinfo{title}{Aerosol and surface stability of {SARS}-{CoV}-2 as
  compared with {SARS}-{CoV}-1},
\newblock \bibinfo{journal}{New England Journal of Medicine}
  \bibinfo{volume}{382} (\bibinfo{year}{2020}) \bibinfo{pages}{1564--1567}.
\bibitem[{Bai et~al.(2020)Bai, Yao, Wei, Tian, Jin, Chen, and
  Wang}]{bai2020presumed}
\bibinfo{author}{Y.~Bai}, \bibinfo{author}{L.~Yao}, \bibinfo{author}{T.~Wei},
  \bibinfo{author}{F.~Tian}, \bibinfo{author}{D.-Y. Jin},
  \bibinfo{author}{L.~Chen}, \bibinfo{author}{M.~Wang},
\newblock \bibinfo{title}{Presumed asymptomatic carrier transmission of
  {COVID}-19},
\newblock \bibinfo{journal}{Jama} \bibinfo{volume}{323} (\bibinfo{year}{2020})
  \bibinfo{pages}{1406--1407}.
\bibitem[{Nishiura et~al.(2020)Nishiura, Kobayashi, Miyama, Suzuki, Jung,
  Hayashi, Kinoshita, Yang, Yuan, Akhmetzhanov et~al.}]{nishiura2020estimation}
\bibinfo{author}{H.~Nishiura}, \bibinfo{author}{T.~Kobayashi},
  \bibinfo{author}{T.~Miyama}, \bibinfo{author}{A.~Suzuki},
  \bibinfo{author}{S.~Jung}, \bibinfo{author}{K.~Hayashi},
  \bibinfo{author}{R.~Kinoshita}, \bibinfo{author}{Y.~Yang},
  \bibinfo{author}{B.~Yuan}, \bibinfo{author}{A.~R. Akhmetzhanov}, et~al.,
\newblock \bibinfo{title}{Estimation of the asymptomatic ratio of novel
  coronavirus infections ({COVID}-19)},
\newblock \bibinfo{journal}{medRxiv}  (\bibinfo{year}{2020}).
\bibitem[{Yu et~al.(2020)Yu, Zhu, Zhang, and Han}]{yu2020familial}
\bibinfo{author}{P.~Yu}, \bibinfo{author}{J.~Zhu}, \bibinfo{author}{Z.~Zhang},
  \bibinfo{author}{Y.~Han},
\newblock \bibinfo{title}{A familial cluster of infection associated with the
  2019 novel coronavirus indicating possible person-to-person transmission
  during the incubation period},
\newblock \bibinfo{journal}{The Journal of infectious diseases}
  \bibinfo{volume}{221} (\bibinfo{year}{2020}) \bibinfo{pages}{1757--1761}.
\bibitem[{Giordano et~al.(2020)Giordano, Blanchini, Bruno, Colaneri,
  Di~Filippo, Di~Matteo, and Colaneri}]{giordano2020modelling}
\bibinfo{author}{G.~Giordano}, \bibinfo{author}{F.~Blanchini},
  \bibinfo{author}{R.~Bruno}, \bibinfo{author}{P.~Colaneri},
  \bibinfo{author}{A.~Di~Filippo}, \bibinfo{author}{A.~Di~Matteo},
  \bibinfo{author}{M.~Colaneri},
\newblock \bibinfo{title}{Modelling the covid-19 epidemic and implementation of
  population-wide interventions in italy},
\newblock \bibinfo{journal}{Nature Medicine}  (\bibinfo{year}{2020})
  \bibinfo{pages}{1--6}.
\bibitem[{Yang et~al.(2020)Yang, Zeng, Wang, Wong, Liang, Zanin, Liu, Cao, Gao,
  Mai et~al.}]{yang2020modified}
\bibinfo{author}{Z.~Yang}, \bibinfo{author}{Z.~Zeng},
  \bibinfo{author}{K.~Wang}, \bibinfo{author}{S.-S. Wong},
  \bibinfo{author}{W.~Liang}, \bibinfo{author}{M.~Zanin},
  \bibinfo{author}{P.~Liu}, \bibinfo{author}{X.~Cao}, \bibinfo{author}{Z.~Gao},
  \bibinfo{author}{Z.~Mai}, et~al.,
\newblock \bibinfo{title}{Modified seir and ai prediction of the epidemics
  trend of covid-19 in china under public health interventions},
\newblock \bibinfo{journal}{Journal of Thoracic Disease} \bibinfo{volume}{12}
  (\bibinfo{year}{2020}) \bibinfo{pages}{165}.
\bibitem[{Volpert et~al.(2020)Volpert, Banerjee, and
  Petrovskii}]{volpert2020quarantine}
\bibinfo{author}{V.~Volpert}, \bibinfo{author}{M.~Banerjee},
  \bibinfo{author}{S.~Petrovskii},
\newblock \bibinfo{title}{On a quarantine model of coronavirus infection and
  data analysis},
\newblock \bibinfo{journal}{Mathematical Modelling of Natural Phenomena}
  \bibinfo{volume}{15} (\bibinfo{year}{2020}) \bibinfo{pages}{24}.
\bibitem[{Li et~al.(2020{\natexlab{a}})Li, Chen, Chen, Zhang, Pang, and
  Chen}]{li2020retrospective}
\bibinfo{author}{C.~Li}, \bibinfo{author}{L.~J. Chen},
  \bibinfo{author}{X.~Chen}, \bibinfo{author}{M.~Zhang}, \bibinfo{author}{C.~P.
  Pang}, \bibinfo{author}{H.~Chen},
\newblock \bibinfo{title}{Retrospective analysis of the possibility of
  predicting the covid-19 outbreak from internet searches and social media
  data, china, 2020},
\newblock \bibinfo{journal}{Eurosurveillance} \bibinfo{volume}{25}
  (\bibinfo{year}{2020}{\natexlab{a}}) \bibinfo{pages}{2000199}.
\bibitem[{Li et~al.(2020{\natexlab{b}})Li, Yang, Dang, Meng, Huang, Meng, Wang,
  Chen, Zhang, Peng et~al.}]{li2020propagation}
\bibinfo{author}{L.~Li}, \bibinfo{author}{Z.~Yang}, \bibinfo{author}{Z.~Dang},
  \bibinfo{author}{C.~Meng}, \bibinfo{author}{J.~Huang},
  \bibinfo{author}{H.~Meng}, \bibinfo{author}{D.~Wang},
  \bibinfo{author}{G.~Chen}, \bibinfo{author}{J.~Zhang},
  \bibinfo{author}{H.~Peng}, et~al.,
\newblock \bibinfo{title}{Propagation analysis and prediction of the covid-19},
\newblock \bibinfo{journal}{Infectious Disease Modelling} \bibinfo{volume}{5}
  (\bibinfo{year}{2020}{\natexlab{b}}) \bibinfo{pages}{282--292}.
\bibitem[{Fong et~al.(2020)Fong, Li, Dey, Crespo, and
  Herrera-Viedma}]{fong2020composite}
\bibinfo{author}{S.~J. Fong}, \bibinfo{author}{G.~Li},
  \bibinfo{author}{N.~Dey}, \bibinfo{author}{R.~G. Crespo},
  \bibinfo{author}{E.~Herrera-Viedma},
\newblock \bibinfo{title}{Composite monte carlo decision making under high
  uncertainty of novel coronavirus epidemic using hybridized deep learning and
  fuzzy rule induction},
\newblock \bibinfo{journal}{Applied Soft Computing}  (\bibinfo{year}{2020})
  \bibinfo{pages}{106282}.
\bibitem[{Chatterjee et~al.(2020)Chatterjee, Chatterjee, Kumar, and
  Shankar}]{chatterjee2020healthcare}
\bibinfo{author}{K.~Chatterjee}, \bibinfo{author}{K.~Chatterjee},
  \bibinfo{author}{A.~Kumar}, \bibinfo{author}{S.~Shankar},
\newblock \bibinfo{title}{Healthcare impact of covid-19 epidemic in india: A
  stochastic mathematical model},
\newblock \bibinfo{journal}{Medical Journal Armed Forces India}
  (\bibinfo{year}{2020}).
\bibitem[{Fong et~al.(2020)Fong, Li, Dey, Crespo, and
  Herrera-Viedma}]{fong2020finding}
\bibinfo{author}{S.~J. Fong}, \bibinfo{author}{G.~Li},
  \bibinfo{author}{N.~Dey}, \bibinfo{author}{R.~G. Crespo},
  \bibinfo{author}{E.~Herrera-Viedma},
\newblock \bibinfo{title}{Finding an accurate early forecasting model from
  small dataset: A case of 2019-ncov novel coronavirus outbreak},
\newblock \bibinfo{journal}{arXiv preprint arXiv:2003.10776}
  (\bibinfo{year}{2020}).
\bibitem[{Baltas et~al.(2020)Baltas, Prieto~Rodr{\'\i}guez, Frantzi,
  Garc{\'\i}a~Alonso, Rodr{\'\i}guez~Cort{\'e}s et~al.}]{baltas2020monte}
\bibinfo{author}{G.~Baltas}, \bibinfo{author}{F.~A. Prieto~Rodr{\'\i}guez},
  \bibinfo{author}{M.~Frantzi}, \bibinfo{author}{C.~Garc{\'\i}a~Alonso},
  \bibinfo{author}{P.~Rodr{\'\i}guez~Cort{\'e}s}, et~al.,
\newblock \bibinfo{title}{Monte carlo deep neural network model for spread and
  peak prediction of covid-19}  (\bibinfo{year}{2020}).
\bibitem[{Khatua et~al.(2020)Khatua, De, Kar, Samanta, Seikh, and
  Guha}]{khatua2020fuzzy}
\bibinfo{author}{D.~Khatua}, \bibinfo{author}{A.~De}, \bibinfo{author}{S.~Kar},
  \bibinfo{author}{E.~Samanta}, \bibinfo{author}{A.~A. Seikh},
  \bibinfo{author}{D.~Guha},
\newblock \bibinfo{title}{A fuzzy dynamic optimal model for covid-19 epidemic
  in india based on granular differentiability},
\newblock \bibinfo{journal}{Available at SSRN 3621640}  (\bibinfo{year}{2020}).
\bibitem[{Liu et~al.(2020)Liu, Beeler, and Chakrabarty}]{liu2020covid}
\bibinfo{author}{P.~Liu}, \bibinfo{author}{P.~Beeler}, \bibinfo{author}{R.~K.
  Chakrabarty},
\newblock \bibinfo{title}{Covid-19 progression timeline and effectiveness of
  response-to-spread interventions across the united states},
\newblock \bibinfo{journal}{medRxiv}  (\bibinfo{year}{2020}).
\bibitem[{Traini et~al.(2020)Traini, Caponi, and
  De~Socio}]{traini2020modelling}
\bibinfo{author}{M.~C. Traini}, \bibinfo{author}{C.~Caponi},
  \bibinfo{author}{G.~V. De~Socio},
\newblock \bibinfo{title}{Modelling the epidemic 2019-ncov event in italy: a
  preliminary note},
\newblock \bibinfo{journal}{medRxiv}  (\bibinfo{year}{2020}).
\bibitem[{Lai et~al.(2020)Lai, Bogoch, Ruktanonchai, Watts, Lu, Yang, Yu, Khan,
  and Tatem}]{lai2020assessing}
\bibinfo{author}{S.~Lai}, \bibinfo{author}{I.~I. Bogoch},
  \bibinfo{author}{N.~W. Ruktanonchai}, \bibinfo{author}{A.~Watts},
  \bibinfo{author}{X.~Lu}, \bibinfo{author}{W.~Yang}, \bibinfo{author}{H.~Yu},
  \bibinfo{author}{K.~Khan}, \bibinfo{author}{A.~J. Tatem},
\newblock \bibinfo{title}{Assessing spread risk of wuhan novel coronavirus
  within and beyond china, january-april 2020: a travel network-based modelling
  study},
\newblock \bibinfo{journal}{medRxiv}  (\bibinfo{year}{2020}).
\bibitem[{Wynants et~al.(2020)Wynants, Van~Calster, Bonten, Collins, Debray,
  De~Vos, Haller, Heinze, Moons, Riley et~al.}]{wynants2020prediction}
\bibinfo{author}{L.~Wynants}, \bibinfo{author}{B.~Van~Calster},
  \bibinfo{author}{M.~M. Bonten}, \bibinfo{author}{G.~S. Collins},
  \bibinfo{author}{T.~P. Debray}, \bibinfo{author}{M.~De~Vos},
  \bibinfo{author}{M.~C. Haller}, \bibinfo{author}{G.~Heinze},
  \bibinfo{author}{K.~G. Moons}, \bibinfo{author}{R.~D. Riley}, et~al.,
\newblock \bibinfo{title}{Prediction models for diagnosis and prognosis of
  covid-19 infection: systematic review and critical appraisal},
\newblock \bibinfo{journal}{bmj} \bibinfo{volume}{369} (\bibinfo{year}{2020}).
\bibitem[{Bauch et~al.(2005)Bauch, Lloyd-Smith, Coffee, and
  Galvani}]{bauch2005dynamically}
\bibinfo{author}{C.~T. Bauch}, \bibinfo{author}{J.~O. Lloyd-Smith},
  \bibinfo{author}{M.~P. Coffee}, \bibinfo{author}{A.~P. Galvani},
\newblock \bibinfo{title}{Dynamically modeling sars and other newly emerging
  respiratory illnesses: past, present, and future},
\newblock \bibinfo{journal}{Epidemiology}  (\bibinfo{year}{2005})
  \bibinfo{pages}{791--801}.
\bibitem[{Shinde et~al.(2020)Shinde, Kalamkar, Mahalle, Dey, Chaki, and
  Hassanien}]{shinde2020forecasting}
\bibinfo{author}{G.~R. Shinde}, \bibinfo{author}{A.~B. Kalamkar},
  \bibinfo{author}{P.~N. Mahalle}, \bibinfo{author}{N.~Dey},
  \bibinfo{author}{J.~Chaki}, \bibinfo{author}{A.~E. Hassanien},
\newblock \bibinfo{title}{Forecasting models for coronavirus disease
  (covid-19): A survey of the state-of-the-art},
\newblock \bibinfo{journal}{SN Computer Science} \bibinfo{volume}{1}
  (\bibinfo{year}{2020}) \bibinfo{pages}{1--15}.
\bibitem[{Althouse et~al.(2012)Althouse, Lessler, Sall, Diallo, Hanley, Watts,
  Weaver, and Cummings}]{althouse2012synchrony}
\bibinfo{author}{B.~M. Althouse}, \bibinfo{author}{J.~Lessler},
  \bibinfo{author}{A.~A. Sall}, \bibinfo{author}{M.~Diallo},
  \bibinfo{author}{K.~A. Hanley}, \bibinfo{author}{D.~M. Watts},
  \bibinfo{author}{S.~C. Weaver}, \bibinfo{author}{D.~A. Cummings},
\newblock \bibinfo{title}{Synchrony of sylvatic dengue isolations: a
  multi-host, multi-vector sir model of dengue virus transmission in senegal},
\newblock \bibinfo{journal}{PLoS Negl Trop Dis} \bibinfo{volume}{6}
  (\bibinfo{year}{2012}) \bibinfo{pages}{e1928}.
\bibitem[{Anderson and May(1992)}]{anderson1992infectious}
\bibinfo{author}{R.~M. Anderson}, \bibinfo{author}{R.~M. May},
  \bibinfo{title}{Infectious diseases of humans: dynamics and control},
  \bibinfo{publisher}{Oxford university press}, \bibinfo{year}{1992}.
\bibitem[{Hethcote(1973)}]{hethcote1973asymptotic}
\bibinfo{author}{H.~W. Hethcote},
\newblock \bibinfo{title}{Asymptotic behavior in a deterministic epidemic
  model},
\newblock \bibinfo{journal}{Bulletin of Mathematical Biology}
  \bibinfo{volume}{35} (\bibinfo{year}{1973}) \bibinfo{pages}{607--614}.
\bibitem[{Behncke(2000)}]{behncke2000optimal}
\bibinfo{author}{H.~Behncke},
\newblock \bibinfo{title}{Optimal control of deterministic epidemics},
\newblock \bibinfo{journal}{Optimal control applications and methods}
  \bibinfo{volume}{21} (\bibinfo{year}{2000}) \bibinfo{pages}{269--285}.
\bibitem[{Bhattacharya et~al.(2019)Bhattacharya, Gaurav, and
  Ghosh}]{bhattacharya2019viral}
\bibinfo{author}{S.~Bhattacharya}, \bibinfo{author}{K.~Gaurav},
  \bibinfo{author}{S.~Ghosh},
\newblock \bibinfo{title}{Viral marketing on social networks: An
  epidemiological perspective},
\newblock \bibinfo{journal}{Physica A: Statistical Mechanics and its
  Applications} \bibinfo{volume}{525} (\bibinfo{year}{2019})
  \bibinfo{pages}{478--490}.
\bibitem[{Liu et~al.(2020)Liu, Gayle, Wilder-Smith, and
  Rockl{\"o}v}]{liu2020reproductive}
\bibinfo{author}{Y.~Liu}, \bibinfo{author}{A.~A. Gayle},
  \bibinfo{author}{A.~Wilder-Smith}, \bibinfo{author}{J.~Rockl{\"o}v},
\newblock \bibinfo{title}{The reproductive number of {COVID}-19 is higher
  compared to {SARS} coronavirus},
\newblock \bibinfo{journal}{Journal of travel medicine}
  (\bibinfo{year}{2020}).
\bibitem[{Shim et~al.(2020)Shim, Tariq, Choi, Lee, and
  Chowell}]{shim2020transmission}
\bibinfo{author}{E.~Shim}, \bibinfo{author}{A.~Tariq},
  \bibinfo{author}{W.~Choi}, \bibinfo{author}{Y.~Lee},
  \bibinfo{author}{G.~Chowell},
\newblock \bibinfo{title}{Transmission potential and severity of {COVID}-19 in
  {S}outh {K}orea},
\newblock \bibinfo{journal}{International Journal of Infectious Diseases}
  (\bibinfo{year}{2020}).
\bibitem[{Kucharski et~al.(2020)Kucharski, Russell, Diamond, Liu, Edmunds,
  Funk, Eggo, Sun, Jit, Munday et~al.}]{kucharski2020early}
\bibinfo{author}{A.~J. Kucharski}, \bibinfo{author}{T.~W. Russell},
  \bibinfo{author}{C.~Diamond}, \bibinfo{author}{Y.~Liu},
  \bibinfo{author}{J.~Edmunds}, \bibinfo{author}{S.~Funk},
  \bibinfo{author}{R.~M. Eggo}, \bibinfo{author}{F.~Sun},
  \bibinfo{author}{M.~Jit}, \bibinfo{author}{J.~D. Munday}, et~al.,
\newblock \bibinfo{title}{Early dynamics of transmission and control of
  covid-19: a mathematical modelling study},
\newblock \bibinfo{journal}{The lancet infectious diseases}
  (\bibinfo{year}{2020}).
\bibitem[{Peng et~al.(2020)Peng, Yang, Zhang, Zhuge, and
  Hong}]{peng2020epidemic}
\bibinfo{author}{L.~Peng}, \bibinfo{author}{W.~Yang},
  \bibinfo{author}{D.~Zhang}, \bibinfo{author}{C.~Zhuge},
  \bibinfo{author}{L.~Hong},
\newblock \bibinfo{title}{Epidemic analysis of covid-19 in china by dynamical
  modeling},
\newblock \bibinfo{journal}{arXiv preprint arXiv:2002.06563}
  (\bibinfo{year}{2020}).
\bibitem[{Kermack and McKendrick(1927)}]{kermack1927contribution}
\bibinfo{author}{W.~O. Kermack}, \bibinfo{author}{A.~G. McKendrick},
\newblock \bibinfo{title}{A contribution to the mathematical theory of
  epidemics},
\newblock \bibinfo{journal}{Proceedings of the royal society of london. Series
  A, Containing papers of a mathematical and physical character}
  \bibinfo{volume}{115} (\bibinfo{year}{1927}) \bibinfo{pages}{700--721}.
\bibitem[{Rachah and Torres(2017)}]{rachah2017analysis}
\bibinfo{author}{A.~Rachah}, \bibinfo{author}{D.~F. Torres},
\newblock \bibinfo{title}{Analysis, simulation and optimal control of a seir
  model for ebola virus with demographic effects},
\newblock \bibinfo{journal}{arXiv preprint arXiv:1705.01079}
  (\bibinfo{year}{2017}).
\bibitem[{Berge et~al.(2017)Berge, Lubuma, Moremedi, Morris, and
  Kondera-Shava}]{berge2017simple}
\bibinfo{author}{T.~Berge}, \bibinfo{author}{J.-S. Lubuma},
  \bibinfo{author}{G.~Moremedi}, \bibinfo{author}{N.~Morris},
  \bibinfo{author}{R.~Kondera-Shava},
\newblock \bibinfo{title}{A simple mathematical model for ebola in africa},
\newblock \bibinfo{journal}{Journal of biological dynamics}
  \bibinfo{volume}{11} (\bibinfo{year}{2017}) \bibinfo{pages}{42--74}.
\bibitem[{Toffoli and Margolus(1987)}]{toffoli1987cellular}
\bibinfo{author}{T.~Toffoli}, \bibinfo{author}{N.~Margolus},
  \bibinfo{title}{Cellular automata machines: a new environment for modeling},
  \bibinfo{publisher}{MIT press}, \bibinfo{year}{1987}.
\bibitem[{Wolfram(2018)}]{wolfram2018cellular}
\bibinfo{author}{S.~Wolfram}, \bibinfo{title}{Cellular automata and complexity:
  collected papers}, \bibinfo{publisher}{CRC Press}, \bibinfo{year}{2018}.
\bibitem[{Boccara et~al.(1994)Boccara, Cheong, and
  Oram}]{boccara1994probabilistic}
\bibinfo{author}{N.~Boccara}, \bibinfo{author}{K.~Cheong},
  \bibinfo{author}{M.~Oram},
\newblock \bibinfo{title}{A probabilistic automata network epidemic model with
  births and deaths exhibiting cyclic behaviour},
\newblock \bibinfo{journal}{Journal of Physics A: Mathematical and General}
  \bibinfo{volume}{27} (\bibinfo{year}{1994}) \bibinfo{pages}{1585}.
\bibitem[{Beauchemin et~al.(2005)Beauchemin, Samuel, and
  Tuszynski}]{beauchemin2005simple}
\bibinfo{author}{C.~Beauchemin}, \bibinfo{author}{J.~Samuel},
  \bibinfo{author}{J.~Tuszynski},
\newblock \bibinfo{title}{A simple cellular automaton model for influenza a
  viral infections},
\newblock \bibinfo{journal}{Journal of theoretical biology}
  \bibinfo{volume}{232} (\bibinfo{year}{2005}) \bibinfo{pages}{223--234}.
\bibitem[{Fuks and Lawniczak(2001)}]{fuks2001individual}
\bibinfo{author}{H.~Fuks}, \bibinfo{author}{A.~T. Lawniczak},
\newblock \bibinfo{title}{Individual-based lattice model for spatial spread of
  epidemics},
\newblock \bibinfo{journal}{Discrete Dynamics in Nature and Society}
  \bibinfo{volume}{6} (\bibinfo{year}{2001}).
\bibitem[{Willox et~al.(2003)Willox, Grammaticos, Carstea, and
  Ramani}]{willox2003epidemic}
\bibinfo{author}{R.~Willox}, \bibinfo{author}{B.~Grammaticos},
  \bibinfo{author}{A.~Carstea}, \bibinfo{author}{A.~Ramani},
\newblock \bibinfo{title}{Epidemic dynamics: discrete-time and cellular
  automaton models},
\newblock \bibinfo{journal}{Physica A: Statistical Mechanics and its
  Applications} \bibinfo{volume}{328} (\bibinfo{year}{2003})
  \bibinfo{pages}{13--22}.
\bibitem[{Eosina et~al.(2016)Eosina, Djatna, and Khusun}]{eosina2016cellular}
\bibinfo{author}{P.~Eosina}, \bibinfo{author}{T.~Djatna},
  \bibinfo{author}{H.~Khusun},
\newblock \bibinfo{title}{A cellular automata modeling for visualizing and
  predicting spreading patterns of dengue fever},
\newblock \bibinfo{journal}{Telkomnika} \bibinfo{volume}{14}
  (\bibinfo{year}{2016}) \bibinfo{pages}{228}.
\bibitem[{Pokkuluri and Nedunuri(2020)}]{pokkuluri2020novel}
\bibinfo{author}{K.~S. Pokkuluri}, \bibinfo{author}{S.~U.~D. Nedunuri},
\newblock \bibinfo{title}{A novel cellular automata classifier for covid-19
  prediction},
\newblock \bibinfo{journal}{Journal of Health Sciences} \bibinfo{volume}{10}
  (\bibinfo{year}{2020}) \bibinfo{pages}{34--38}.
\bibitem[{Dascalu et~al.(2020)Dascalu, Malita, Barbilian, Franti, and
  Stefan}]{dascalu2020enhanced}
\bibinfo{author}{M.~Dascalu}, \bibinfo{author}{M.~Malita},
  \bibinfo{author}{A.~Barbilian}, \bibinfo{author}{E.~Franti},
  \bibinfo{author}{G.~M. Stefan},
\newblock \bibinfo{title}{Enhanced cellular automata with autonomous agents for
  covid-19 pandemic modeling},
\newblock \bibinfo{journal}{ROMANIAN JOURNAL OF INFORMATION SCIENCE AND
  TECHNOLOGY} \bibinfo{volume}{23} (\bibinfo{year}{2020})
  \bibinfo{pages}{S15--S27}.
\bibitem[{Ghosh and Bhattacharya(2020)}]{ghosh2020computational}
\bibinfo{author}{S.~Ghosh}, \bibinfo{author}{S.~Bhattacharya},
\newblock \bibinfo{title}{Computational model on covid-19 pandemic using
  probabilistic cellular automata},
\newblock \bibinfo{journal}{arXiv preprint arXiv:2006.11270}
  (\bibinfo{year}{2020}).
\bibitem[{Wright(1991)}]{wright1991genetic}
\bibinfo{author}{A.~H. Wright},
\newblock \bibinfo{title}{Genetic algorithms for real parameter optimization},
\newblock in: \bibinfo{booktitle}{Foundations of genetic algorithms},
  volume~\bibinfo{volume}{1}, \bibinfo{publisher}{Elsevier},
  \bibinfo{year}{1991}, pp. \bibinfo{pages}{205--218}.
\bibitem[{Yao and Sethares(1994)}]{yao1994nonlinear}
\bibinfo{author}{L.~Yao}, \bibinfo{author}{W.~A. Sethares},
\newblock \bibinfo{title}{Nonlinear parameter estimation via the genetic
  algorithm},
\newblock \bibinfo{journal}{IEEE Transactions on signal processing}
  \bibinfo{volume}{42} (\bibinfo{year}{1994}) \bibinfo{pages}{927--935}.
\bibitem[{Katare et~al.(2004)Katare, Bhan, Caruthers, Delgass, and
  Venkatasubramanian}]{katare2004hybrid}
\bibinfo{author}{S.~Katare}, \bibinfo{author}{A.~Bhan}, \bibinfo{author}{J.~M.
  Caruthers}, \bibinfo{author}{W.~N. Delgass},
  \bibinfo{author}{V.~Venkatasubramanian},
\newblock \bibinfo{title}{A hybrid genetic algorithm for efficient parameter
  estimation of large kinetic models},
\newblock \bibinfo{journal}{Computers \& chemical engineering}
  \bibinfo{volume}{28} (\bibinfo{year}{2004}) \bibinfo{pages}{2569--2581}.
\bibitem[{Gulsen et~al.(1995)Gulsen, Smith, and Tate}]{gulsen1995genetic}
\bibinfo{author}{M.~Gulsen}, \bibinfo{author}{A.~Smith},
  \bibinfo{author}{D.~Tate},
\newblock \bibinfo{title}{A genetic algorithm approach to curve fitting},
\newblock \bibinfo{journal}{International Journal of Production Research}
  \bibinfo{volume}{33} (\bibinfo{year}{1995}) \bibinfo{pages}{1911--1923}.
\bibitem[{Karr et~al.(1995)Karr, Weck, Massart, and
  Vankeerberghen}]{karr1995least}
\bibinfo{author}{C.~L. Karr}, \bibinfo{author}{B.~Weck}, \bibinfo{author}{D.-L.
  Massart}, \bibinfo{author}{P.~Vankeerberghen},
\newblock \bibinfo{title}{Least median squares curve fitting using a genetic
  algorithm},
\newblock \bibinfo{journal}{Engineering Applications of Artificial
  Intelligence} \bibinfo{volume}{8} (\bibinfo{year}{1995})
  \bibinfo{pages}{177--189}.
\bibitem[{Schimit(2016)}]{schimit2016evolutionary}
\bibinfo{author}{P.~H. Schimit},
\newblock \bibinfo{title}{Evolutionary aspects of spatial prisoner’s dilemma
  in a population modeled by continuous probabilistic cellular automata and
  genetic algorithm.},
\newblock \bibinfo{journal}{Applied Mathematics and Computation}
  \bibinfo{volume}{290} (\bibinfo{year}{2016}) \bibinfo{pages}{178--188}.
\bibitem[{Holland et~al.(1992)}]{holland1992adaptation}
\bibinfo{author}{J.~H. Holland}, et~al., \bibinfo{title}{Adaptation in natural
  and artificial systems: an introductory analysis with applications to
  biology, control, and artificial intelligence}, \bibinfo{publisher}{MIT
  press}, \bibinfo{year}{1992}.
\bibitem[{Liao(2005)}]{liao2005clustering}
\bibinfo{author}{T.~W. Liao},
\newblock \bibinfo{title}{Clustering of time series data—a survey},
\newblock \bibinfo{journal}{Pattern recognition} \bibinfo{volume}{38}
  (\bibinfo{year}{2005}) \bibinfo{pages}{1857--1874}.
\bibitem[{Gao et~al.(2009)Gao, Sultan, Hu, and Tung}]{gao2009denoising}
\bibinfo{author}{J.~Gao}, \bibinfo{author}{H.~Sultan}, \bibinfo{author}{J.~Hu},
  \bibinfo{author}{W.-W. Tung},
\newblock \bibinfo{title}{Denoising nonlinear time series by adaptive filtering
  and wavelet shrinkage: a comparison},
\newblock \bibinfo{journal}{IEEE signal processing letters}
  \bibinfo{volume}{17} (\bibinfo{year}{2009}) \bibinfo{pages}{237--240}.
\bibitem[{Salem et~al.(2014)Salem, Liu, and Mehaoua}]{salem2014anomaly}
\bibinfo{author}{O.~Salem}, \bibinfo{author}{Y.~Liu},
  \bibinfo{author}{A.~Mehaoua},
\newblock \bibinfo{title}{Anomaly detection in medical wsns using enclosing
  ellipse and chi-square distance},
\newblock in: \bibinfo{booktitle}{2014 IEEE International Conference on
  Communications (ICC)}, \bibinfo{organization}{IEEE}, pp.
  \bibinfo{pages}{3658--3663}.
\bibitem[{{\relax{World Health Organization coronavirus disease (COVID-2019)
  situation reports}}(2020)}]{who2020incubation}
\bibinfo{author}{{\relax{World Health Organization coronavirus disease
  (COVID-2019) situation reports}}}, \bibinfo{title}{Available at url:
  \url{https://www.who.int/docs/default-source/coronaviruse/situation-reports/20200402-sitrep-73-covid-19.pdf}
  (accessed {J}une 2020)}, \bibinfo{year}{2020}.
\bibitem[{Acharjya and Anitha(2017)}]{acharjya2017comparative}
\bibinfo{author}{D.~Acharjya}, \bibinfo{author}{A.~Anitha},
\newblock \bibinfo{title}{A comparative study of statistical and rough
  computing models in predictive data analysis},
\newblock \bibinfo{journal}{International Journal of Ambient Computing and
  Intelligence (IJACI)} \bibinfo{volume}{8} (\bibinfo{year}{2017})
  \bibinfo{pages}{32--51}.
\bibitem[{Su and Zhang(2006)}]{su2006fast}
\bibinfo{author}{J.~Su}, \bibinfo{author}{H.~Zhang},
\newblock \bibinfo{title}{A fast decision tree learning algorithm},
\newblock in: \bibinfo{booktitle}{AAAI}, volume~\bibinfo{volume}{6}, pp.
  \bibinfo{pages}{500--505}.
\bibitem[{Miller et~al.(2020)Miller, Reandelar, Fasciglione, Roumenova, Li, and
  Otazu}]{miller2020correlation}
\bibinfo{author}{A.~Miller}, \bibinfo{author}{M.~J. Reandelar},
  \bibinfo{author}{K.~Fasciglione}, \bibinfo{author}{V.~Roumenova},
  \bibinfo{author}{Y.~Li}, \bibinfo{author}{G.~H. Otazu},
\newblock \bibinfo{title}{Correlation between universal bcg vaccination policy
  and reduced morbidity and mortality for covid-19: an epidemiological study},
\newblock \bibinfo{journal}{MedRxiv}  (\bibinfo{year}{2020}).

\end{thebibliography}
\end{document}